\documentclass[aps,twocolumn,amsmath,amssymb,floatfix,superscriptaddress]{revtex4-1}

\usepackage{graphicx}
\usepackage{color}
\usepackage{hyperref}

\begin{document}

\title{Bursty time series analysis for temporal networks}\thanks{This is a preprint for a chapter to appear in \emph{Temporal Network Theory} edited by P. Holme and J. Saram\"aki (\url{https://www.springer.com/gp/book/9783030234942}).}

\author{Hang-Hyun Jo}
\email{hang-hyun.jo@apctp.org}
\affiliation{Asia Pacific Center for Theoretical Physics, Pohang 37673, Republic of Korea}
\affiliation{Department of Physics, Pohang University of Science and Technology, Pohang 37673, Republic of Korea}
\affiliation{Department of Computer Science, Aalto University, Espoo FI-00076, Finland}
\author{Takayuki Hiraoka}
\affiliation{Asia Pacific Center for Theoretical Physics, Pohang 37673, Republic of Korea}

\date{\today}

\begin{abstract}
  Characterizing bursty temporal interaction patterns of temporal networks is crucial to investigate the evolution of temporal networks as well as various collective dynamics taking place in them. The temporal interaction patterns have been described by a series of interaction events or event sequences, often showing non-Poissonian or bursty nature. Such bursty event sequences can be understood not only by heterogeneous interevent times (IETs) but also by correlations between IETs. The heterogeneities of IETs have been extensively studied in recent years, while the correlations between IETs are far from being fully explored. In this Chapter, we introduce various measures for bursty time series analysis, such as the IET distribution, the burstiness parameter, the memory coefficient, the bursty train sizes, and the autocorrelation function, to discuss the relation between those measures. Then we show that the correlations between IETs can affect the speed of spreading taking place in temporal networks. Finally, we discuss possible research topics regarding bursty time series analysis for temporal networks.
\end{abstract}

\maketitle

\section{Introduction}\label{sect:intro}

Characterizing the interaction structure between constituents of complex systems is crucial to understand not only the dynamics of those systems but also the dynamical processes taking place in them. The topological structure of interaction has been modeled by a network, where nodes and links denote the constituents and their pairwise interactions, respectively~\cite{Albert2002Statistical, Newman2010Networks}. When the interaction is temporal, one can adopt a framework of temporal networks~\cite{Holme2012Temporal, Masuda2016Guide, Gauvin2018Randomized}, where links are considered being existent or activated only at the moment of interaction. The temporal interaction pattern of each link can be described by a series of interaction events or an event sequence. Many empirical event sequences are known to be non-Poissonian or bursty~\cite{Barabasi2005Origin, Karsai2012Universal, Karsai2018Bursty}, e.g., as shown in human communication patterns~\cite{Eckmann2004Entropy, Malmgren2009Universality, Cattuto2010Dynamics, Jo2012Circadian, Rybski2012Communication, Jiang2013Calling, Stopczynski2014Measuring, Panzarasa2015Emergence}, where bursts denote a number of events occurring in short active periods separated by long inactive periods. Such bursty event sequences can be fully understood both by heterogeneous interevent times (IETs) and by correlations between IETs~\cite{Goh2008Burstiness, Jo2017Modeling}. Here the IET, denoted by $\tau$, is defined by the time interval between two consecutive IETs. The heterogeneities of IETs have been extensively studied in terms of heavy-tailed or power-law IET distributions~\cite{Karsai2018Bursty}, while the correlations between IETs have been far from being fully explored.

In this Chapter, we introduce various measures for bursty time series analysis, such as the IET distribution, the burstiness parameter, the memory coefficient, the bursty train sizes, and the autocorrelation function, to discuss the relation between those measures. Then we show that the correlations between IETs can affect the speed of spreading taking place in temporal networks. Finally, we discuss possible research topics regarding bursty time series analysis for temporal networks.

\section{Bursty time series analysis}\label{sec:analysis}

\subsection{Measures and characterizations}\label{subsec:measure}

\begin{figure}[!t]
    \includegraphics[width=\columnwidth]{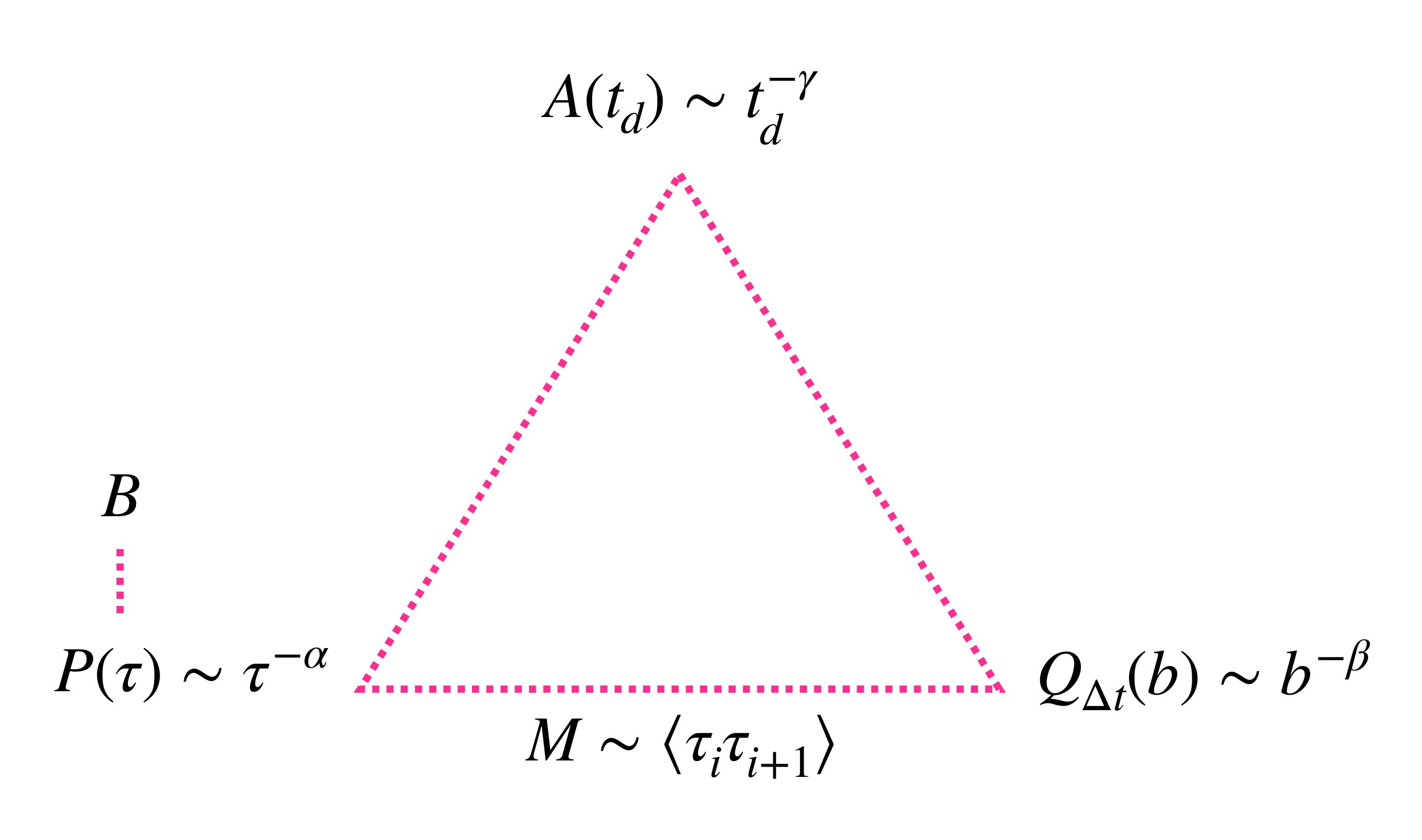}
    \caption{Conceptual diagram for the relation between the autocorrelation function $A(t_d)$, the interevent time distribution $P(\tau)$, and the burst size distribution for a given time window $Q_{\Delta t}(b)$, together with the burstiness parameter $B$ and the memory coefficient $M$. The relation between these five measures is discussed in Section~\ref{sec:analysis}: In particular, for the dependence of $\gamma$ on $\alpha$ and $\beta$, refer to Subsection~\ref{subsec:scaling}, and for the relation between $M$ and $Q_{\Delta t}(b)$, refer to Subsection~\ref{subsec:M_beta}.}
    \label{fig:measures}
\end{figure}

Non-Poissonian, bursty time series or event sequences have been observed not only in the human communication patterns~\cite{Karsai2018Bursty}, but also in other natural and biological phenomena, including solar flares~\cite{Wheatland1998WaitingTime}, earthquakes~\cite{Corral2004LongTerm, deArcangelis2006Universality}, neuronal firings~\cite{Kemuriyama2010Powerlaw}, and animal behaviors~\cite{Sorribes2011Origin, Boyer2012Nonrandom}. Temporal correlations in such event sequences have been characterized by various measures and quantities~\cite{Karsai2018Bursty}, such as the IET distribution, the burstiness parameter, the memory coefficient, the bursty train sizes, and the autocorrelation function. Each of these five measures captures a different aspect of the bursty time series, while they are not independent of each other. Here we discuss the relation between these five measures, which is conceptually illustrated in Fig.~\ref{fig:measures}.

(i) The autocorrelation function for an event sequence $x(t)$ is defined with delay time $t_d$ as follows:
\begin{equation}
  A(t_d)\equiv \frac{ \langle x(t)x(t+t_d)\rangle_t- \langle x(t)\rangle^2_t}{ \langle x(t)^2\rangle_t- \langle x(t)\rangle^2_t},
\end{equation}
where $\langle\cdot\rangle_t$ means a time average. The event sequence $x(t)$ can be considered to have the value of $1$ at the moment of event occurred, $0$ otherwise. For the event sequences with long-term memory effects, one may find a power-law decaying behavior with a decaying exponent $\gamma$:
\begin{equation}
    A(t_d)\sim t_d^{-\gamma}.
\end{equation}
Temporal correlations measured by $A(t_d)$ can be understood not only by the heterogeneous IETs but also by correlations between them.

(ii) The heterogeneous properties of IETs have often been characterized by the heavy-tailed or power-law IET distribution $P(\tau)$ with a power-law exponent $\alpha$:
\begin{equation}
    \label{eq:Ptau_simple}
    P(\tau)\sim \tau^{-\alpha},
\end{equation}
which may already imply clustered short IETs even with no correlations between IETs. In the case when IETs are fully uncorrelated with each other, i.e., for renewal processes~\cite{Mainardi2007Poisson}, the power spectral density was analytically calculated from power-law IET distributions~\cite{Lowen1993Fractal}. Using this result, one can straightforwardly derive the scaling relation between $\alpha$ and $\gamma$:
\begin{eqnarray}
    \label{eq:alpha_gamma}
    \begin{tabular}{ll}
        $\alpha+\gamma=2$ & for $1<\alpha\leq 2$,\\
        $\alpha-\gamma=2$ & for $2<\alpha\leq 3$.
    \end{tabular}
\end{eqnarray}
This relation was also derived in the study of priority queueing models~\cite{Vajna2013Modelling}. The relation $\alpha+\gamma=2$ for $1<\alpha\leq 2$ has been derived in the context of earthquakes~\cite{Abe2009Violation} as well as of the hierarchical burst model~\cite{Lee2018Hierarchical}.

(iii) The degree of burstiness in the event sequence can be measured by a single value derived from the IET distribution, namely, the burstiness parameter $B$, which is defined as~\cite{Goh2008Burstiness}
\begin{equation}
    B \equiv \frac{\sigma - \langle\tau\rangle}{\sigma + \langle\tau\rangle},
\end{equation}
where $\sigma$ and $\langle\tau\rangle$ are the standard deviation and mean of IETs, respectively. For the regular event sequence, all IETs are the same, leading to $B=-1$, while for the totally random, Poisson process, since $\sigma=\langle\tau\rangle$, one gets $B=0$. In the extremely bursty case, characterized by $\sigma\gg \langle\tau\rangle$, one finds $B\to 1$. However, when analyzing the empirical event sequences of finite sizes, the value of $\sigma$ is typically limited by the number of events $n$ such that the maximum value of $\sigma$ turns out to be $\sigma_{\rm max}\simeq \langle\tau\rangle\sqrt{n-1}$, allowing to propose an alternative burstiness measure~\cite{Kim2016Measuring}:
\begin{equation}
    B_n \equiv \frac{\sqrt{n+1}\sigma - \sqrt{n-1}\langle\tau\rangle}{(\sqrt{n+1}-2)\sigma + \sqrt{n-1}\langle\tau\rangle},
\end{equation}
which can have the value of $1$ ($0$) in the extremely bursty case (in the Poisson process) for any $n$.

(iv) The correlations between IETs have been characterized by several measures~\cite{Karsai2018Bursty}. Among them, we focus on the memory coefficient and bursty train sizes. The memory coefficient $M$ is defined as the Pearson correlation coefficient between two consecutive IETs, whose value for a sequence of $n$ IETs, i.e., $\{\tau_i\}_{i=1,\cdots,n}$, can be estimated by~\cite{Goh2008Burstiness} 
\begin{equation}
    M \equiv\frac{1}{n - 1}\sum_{i=1}^{n-1}\frac{(\tau_i - \langle\tau\rangle_1)(\tau_{i+1} - \langle\tau\rangle_2)}{\sigma_1 \sigma_2},
    \label{eq:memory_original}
\end{equation}
where $\langle\tau\rangle_1$ ($\langle\tau\rangle_2$) and $\sigma_1$ ($\sigma_2$) are the average and the standard deviation of the first (last) $n-1$ IETs, respectively. Positive $M$ implies that the small (large) IETs tend to be followed by small (large) IETs. Negative $M$ implies the opposite tendency, while $M=0$ is for the uncorrelated IETs. In many empirical analyses, positive $M$ has been observed~\cite{Goh2008Burstiness, Wang2015Temporal, Guo2017Bounds, Bottcher2017Temporal}. 

(v) Another notion for measuring the correlations between IETs is the bursty trains~\cite{Karsai2012Universal}. A bursty train is defined as a set of consecutive events such that IETs between any two consecutive events in the bursty train are less than or equal to a given time window $\Delta t$, while those between events in different bursty trains are larger than $\Delta t$. The number of events in the bursty train is called bursty train size or burst size, and it is denoted by $b$. The distribution of $b$ would follow an exponential function if the IETs are fully uncorrelated with each other. However, $b$ has been empirically found to be power-law distributed, i.e.,
\begin{equation}
    \label{eq:burstSizeDistribution}
    Q_{\Delta t}(b)\sim b^{-\beta} 
\end{equation}
for a wide range of $\Delta t$, e.g., in earthquakes, neuronal activities, and human communication patterns~\cite{Karsai2012Universal, Karsai2012Correlated, Yasseri2012Dynamics, Wang2015Temporal}. This indicates the presence of higher-order correlations between IETs beyond the correlations measured by $M$\footnote{The generalized memory coefficient has also been suggested as the Pearson correlation coefficient between two IETs separated by $k$ IETs~\cite{Goh2008Burstiness}. The case with $k=0$ corresponds to the $M$ in Eq.~(\ref{eq:memory_original}). The relation between the generalized memory coefficients and burst size distributions can be studied for better understanding the correlation structure between IETs.}. We note that the exponential distributions of $Q_{\Delta t}(b)$ have also been reported for mobile phone calls of individual users in another work~\cite{Jiang2016Twostate}. 

We show that the statistics of IETs and burst sizes are interrelated to each other~\cite{Jo2017Modeling}. Let us consider an event sequence with $n+1$ events and $n$ IETs, denoted by $\mathcal{T}\equiv\{\tau_1,\cdots, \tau_n\}$. For a given $\Delta t$ one can detect $m$ bursty trains whose sizes are denoted by $\mathcal{B}\equiv\{b_1,\cdots, b_m\}$. The sum of burst sizes must be the number of events, i.e., $\sum_{j=1}^m b_j=n+1$. With $\langle b\rangle$ denoting the average burst size, we can write
\begin{equation}
    \label{eq:mn_event}
    m\langle b\rangle = n+1 \simeq n,
\end{equation}
where $n\gg 1$ is assumed. The number of bursty trains is related to the number of IETs larger than $\Delta t$, i.e., 
\begin{equation}
    \label{eq:mn_condition}
    m=|\{\tau_i |\tau_i>\Delta t\}|+1. 
\end{equation}
It is because each burst size, say $b$, requires $b-1$ consecutive IETs less than or equal to $\Delta t$ and one IET larger than $\Delta t$.
In the case with $n,m\gg 1$, we get 
\begin{equation}
    \label{eq:mn_burst}
    m\simeq n F(\Delta t),
\end{equation}
where $F(\Delta t)\equiv \int_{\Delta t}^\infty P(\tau')d\tau'$ denotes the complementary cumulative distribution function of $P(\tau)$. By combining Eqs.~(\ref{eq:mn_event}) and (\ref{eq:mn_burst}), we obtain a general relation as
\begin{equation}
    \label{eq:mn_event_burst}
    \langle b\rangle F(\Delta t) \simeq 1,
\end{equation}
which holds for arbitrary functional forms of $P(\tau)$ and $Q_{\Delta t}(b)$~\cite{Jo2017Modeling}.

\subsection{Correlation structure and the bursty-get-burstier mechanism}\label{subsec:structure}

\begin{figure}[!t]
    \includegraphics[width=\columnwidth]{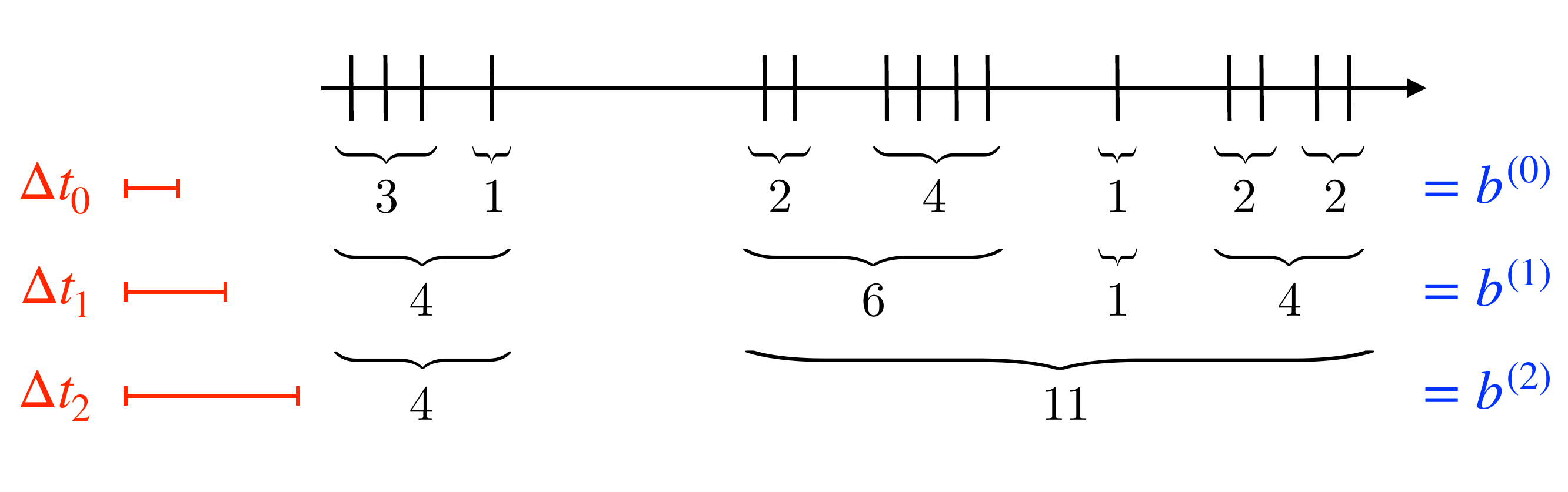}
    \caption{Schematic diagram for the hierarchical organization of bursty trains at various timescales with $15$ events, denoted by vertical lines. These events are clustered using time windows $\Delta t_l$ with $l=0,1,2$, and the sizes of bursty trains or burst sizes are denoted by $b^{(l)}$, e.g., $\{b^{(1)}\}=\{4,6,1,4\}$.}
    \label{fig:correl_structure}
\end{figure}

We pay special attention to the empirical observation that the tail parts of burst size distributions are characterized by the same power-law exponent for a wide range of time windows, e.g., ranging from a few minutes to the order of one hour in mobile phone communication patterns~\cite{Karsai2012Universal}. To better understand this observation, let us begin with a simple example in Fig.~\ref{fig:correl_structure}. For each given time window $\Delta t_l$ with ``level'' $l=0,1,2$, one can obtain the corresponding set of burst sizes, denoted by $\{b^{(l)}\}$. Here we observe that several bursty trains at the level $l$ are merged to make one bursty train at the level $l+1$. In other words, one burst size in $\{b^{(l+1)}\}$ is typically written as a sum of several burst sizes in $\{b^{(l)}\}$. By characterizing this merging pattern one can get insight into the correlation structure between IETs. In particular, we raise a question: In order to find the power-law tail as $Q_{\Delta t_l}(b^{(l)})\sim b^{(l)-\beta}$ for every $l$, which burst sizes in $\{b^{(l)}\}$ should be merged to make one burst size in $\{b^{(l+1)}\}$? One possible answer to this question has recently been suggested, which is called the bursty-get-burstier (BGB) mechanism~\cite{Jo2017Modeling}, indicating that the bigger (smaller) bursty trains tend to follow the bigger (smaller) ones. 

We introduce one implementation method of the BGB mechanism following Ref.~\cite{Jo2017Modeling}, where $P(\tau)$ and $Q_{\Delta t_0}(b^{(0)})$ are assumed to be given. Although this method has been suggested for arbitrary forms of $P(\tau)$ and $Q_{\Delta t_0}(b^{(0)})$, we focus on the case with power-law tails for both distributions. Precisely, we consider a power-law $P(\tau)$ with a power-law exponent $\alpha>1$ and a lower bound of IET $\tau_{\rm min}$, i.e.,
\begin{equation}
    \label{eq:Ptau}
    P(\tau)= (\alpha-1)\tau_{\rm min}^{\alpha-1}\tau^{-\alpha}\theta(\tau-\tau_{\rm min}),
\end{equation}
and a power-law distribution of burst sizes at the zeroth level ($l=0$) as
\begin{equation}
    \label{eq:Qb0}
    Q_{\Delta t_0}(b^{(0)}) = \zeta(\beta)^{-1} b^{(0)-\beta}\ \textrm{for}\ b^{(0)}=1,2,\cdots,
\end{equation}
where $\theta(\cdot)$ denotes the Heaviside step function and $\zeta(\cdot)$ does the Riemann zeta function. 

We first prepare a set of $n$ IETs, $\mathcal{T}= \{\tau_1,\cdots,\tau_n\}$, that are independently drawn from $P(\tau)$ in Eq.~(\ref{eq:Ptau}). This $\mathcal{T}$ is partitioned into several subsets, denoted by $\mathcal{T}_l$, at different timescales or levels $l=0,1,\cdots,L$:
\begin{eqnarray}
    \mathcal{T}_0 &\equiv & \{\tau_i| \tau_{\rm min} \leq \tau_i \leq \Delta t_0\}, \nonumber\\
    \mathcal{T}_l &\equiv & \{\tau_i| \Delta t_{l -1} < \tau_i \leq \Delta t_l \}\ \textrm{for}\ l=1,\cdots,L-1, \\
    \mathcal{T}_L &\equiv & \{\tau_i| \tau_i > \Delta t_{L-1} \},\nonumber
\end{eqnarray}
where $\Delta t_l < \Delta t_{l+1}$ for all $l$s. For example, one can use $\Delta t_l= \tau_{\rm min} cs^l$ with constants $c,s > 1$. This partition readily determines the number of bursty trains at each level, denoted by $m_l$, similarly to Eq.~(\ref{eq:mn_condition}):
\begin{equation}
    m_l = \big|\{\tau_i| \tau_i > \Delta t_l\}\big|+1.
\end{equation}
Then the sizes of bursty trains for a given $\Delta t_l$ are denoted by $\mathcal{B}_l\equiv \{b^{(l)}\}$, with $m_l=|\mathcal{B}_l|$. To generate $\mathcal{B}_0$, $m_0$ burst sizes are independently drawn from $Q_{\Delta t_0}(b^{(0)})$ in Eq.~(\ref{eq:Qb0}). Partitioning $\mathcal{B}_0$ into subsets and summing up the burst sizes in each subset leads to $\mathcal{B}_1$. Precisely, for each $l$, $\mathcal{B}_l$ is sorted, e.g., in a descending order, then it is sequentially partitioned into $m_{l+1}$ subsets of the (almost) same size. The sum of $b^{(l)}$s in each subset leads to one $b^{(l+1)}$, implying that the bigger bursty trains are merged together, so do the smaller ones. This procedure is repeated until the level $L$ is reached. Using the information on which burst sizes at the level $l$ are merged to get each of burst sizes at the level $l+1$, one can construct the sequence of IETs by permuting IETs in $\mathcal{T}$ and finally get the event sequence. See Ref.~\cite{Jo2017Modeling} for details. Numerical simulations have shown that the generated event sequences show $Q_{\Delta t_l}(b^{(l)})\sim b^{(l)-\beta}$ at all levels~\cite{Jo2017Modeling}.

We remark that the above method lacks some realistic features observed in the empirical analyses. For example, by the above method the number of burst sizes in each partition at the level $l$ is almost the same as being either $\lfloor \frac{m_{l}}{m_{l+1}}\rfloor$ or $\lfloor \frac{m_{l}}{m_{l+1}}\rfloor+1$, which is not always the case. Therefore more realistic merging processes for the correlation structure between IETs could be investigated as a future work.

\subsection{Temporal scaling behaviors}\label{subsec:scaling}

The scaling relation between $\alpha$ and $\gamma$ for the uncorrelated IETs in Eq.~(\ref{eq:alpha_gamma}) implies that the autocorrelation function is solely determined by the IET distribution. We can consider a more general case that the IETs are correlated with each other, in particular, in terms of the power-law burst size distributions. Then the temporal correlations measured by the autocorrelation function $A(t_d)$ can be understood by means of the statistical properties of IETs, $P(\tau)$, together with those of the correlations between IETs, $Q_{\Delta t}(b)$. In terms of scaling behaviors, one can study the dependence of $\gamma$ on $\alpha$ and $\beta$.

The dependence of $\gamma$ on $\alpha$ and $\beta$ has been investigated by dynamically generating event sequences showing temporal correlations, described by the power-law distributions of IETs and burst sizes in Eqs.~(\ref{eq:Ptau_simple}) and~(\ref{eq:burstSizeDistribution}). These generative approaches have been based on two-state Markov chain~\cite{Karsai2012Universal} or self-exciting point processes~\cite{Jo2015Correlated}. One can also take an alternative approach by shuffling or permuting a given set of IETs according to the BGB mechanism described in Subsection~\ref{subsec:structure}, where power-law distributions of IETs and burst sizes are inputs rather than outputs of the model. Then one can explicitly tune the degree of correlations between IETs to test whether the scaling relation in Eq.~(\ref{eq:alpha_gamma}) will be violated due to the correlations between IETs. 

\begin{figure}[!t]
    \includegraphics[width=\columnwidth]{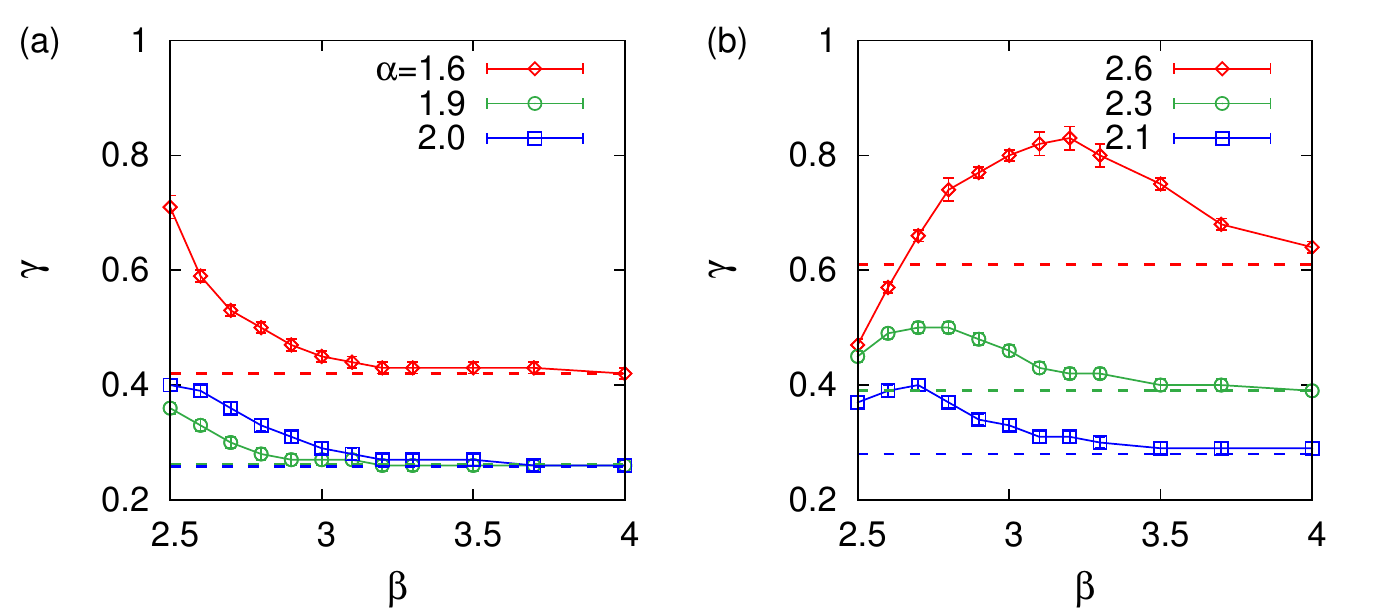}
    \caption{The values of $\gamma$ estimated from the numerically obtained autocorrelation functions, for various values of $\alpha$ and $\beta$, with horizontal dashed lines corresponding to those for the uncorrelated cases. Reprinted figure with permission from Ref.~\cite{Jo2017Modeling} Copyright (2017) by the American Physical Society.}
    \label{fig:correlated_gamma}
\end{figure}

For this, the event sequences are generated using the BGB mechanism for the power-law distributions of IETs and burst sizes, which are then analyzed by measuring autocorrelation functions $A(t_d)$ for various values of $\alpha$ and $\beta$. The decaying exponent $\gamma$ of $A(t_d)$ is estimated based on the simple scaling form of $A(t_d)\sim t_d^{-\gamma}$. The estimated values of $\gamma$ for various values of $\alpha$ and $\beta$ are presented in Fig.~\ref{fig:correlated_gamma}. When $\alpha\leq 2$, it is numerically found that the autocorrelation functions for $\beta < 3$ deviate from the uncorrelated case, implying the violation of scaling relation between $\alpha$ and $\gamma$ in Eq.~(\ref{eq:alpha_gamma}). Precisely, the smaller $\beta$ leads to the larger $\gamma$, implying that the stronger correlations between IETs may induce the faster decaying of autocorrelation. On the other hand, in the case with $\alpha>2$, the estimated $\gamma$ deviates significantly from that for the uncorrelated case for the almost entire range of $\beta$, although $\gamma$ approaches the uncorrelated case as $\beta$ increases as expected.

One can argue that the deviation (or the violation of $\alpha+\gamma=2$) observed for $\beta<3$ is due to the fact that the variance of $b$ diverges for $\beta<3$. This argument seems to explain why $\alpha+\gamma=2$ is observed even when $\beta=3$, for event sequences generated using two-state Markov chain~\cite{Karsai2012Universal}. 

For better understanding the above results, more rigorous studies need to be done. As the analytical calculation of $\gamma$ as a function of $\beta$ is a very challenging task, one can tackle a simplified problem. For example, the effects of correlations only between two consecutive IETs on the autocorrelation function have been analytically studied to find the $M$-dependence of $\gamma$~\cite{Jo2019Analytically}.

\subsection{Limits of the memory coefficient in measuring correlations}\label{subsec:M_beta}

The memory coefficient, measuring the correlations only between two consecutive IETs, has been used to analyze event sequences in natural phenomena and human activities as well as to test models for bursty dynamics~\cite{Goh2008Burstiness, Wang2015Temporal, Bottcher2017Temporal, Jo2015Correlated}. It has been found that $M\approx 0.2$ for earthquakes in Japan, while $M$ is close to $0$ or less than $0.1$ for various human activities~\cite{Goh2008Burstiness}. In another work on emergency call records in a Chinese city, individual callers are found to show diverse values of $M$, i.e., a broad distribution of $M$ ranging from $-0.2$ to $0.5$ but peaked at $M=0$~\cite{Wang2015Temporal}. Based on these empirical observations, one might conclude that most of human activities do not show strong correlations between IETs. On the other hand, the empirical value of $\beta$ for the burst size distributions varies from $2.5$ for earthquakes in Japan to $2.8$--$3.0$ for Wikipedia editing patterns~\cite{Yasseri2012Dynamics} and $3.9$--$4.2$ for mobile phone communication patterns~\cite{Karsai2012Universal, Karsai2012Correlated}, while it is found that $\beta\approx 2.21$ in the emergency call dataset~\cite{Wang2015Temporal}. Since the power-law behaviors of burst size distributions for a wide range of time windows imply the complex, higher-order correlations between IETs, this seems to be inconsistent with the weak correlation implied by the observation $M\approx 0$ in human activities. 

This puzzling issue has been resolved by deriving the analytical form of $M$ as a function of parameters describing $P(\tau)$ and $Q_{\Delta t}(b)$~\cite{Jo2018Limits}. Here we introduce the derivation of $M$ following Ref.~\cite{Jo2018Limits}. By considering bursty trains detected using one time window or timescale $\Delta t$, we divide $\mathcal{T}=\{\tau_1,\cdots, \tau_n\}$ into two subsets as
\begin{eqnarray}
    \label{eq:T_subsets}
    \mathcal{T}_0 &\equiv & \{\tau_i| \tau_i\leq \Delta t\},\\
    \mathcal{T}_1 &\equiv & \{\tau_i| \tau_i> \Delta t\}.
\end{eqnarray}
The set of all pairs of two consecutive IETs, $\{(\tau_i,\tau_{i+1})\}$, can be divided into four subsets as follows:
\begin{equation}
    \label{eq:subsets_pair}
    \mathcal{T}_{\mu\nu} \equiv \{(\tau_i,\tau_{i+1})|\tau_i\in \mathcal{T}_\mu, \tau_{i+1}\in \mathcal{T}_\nu\},
\end{equation}
where $\mu,\nu \in \{0,1\}$. By denoting the fraction of IET pairs in each $\mathcal{T}_{\mu\nu}$ by $t_{\mu\nu}\equiv \langle |\mathcal{T}_{\mu\nu}|\rangle / (n-1)$, the term $\langle\tau_i\tau_{i+1}\rangle$ in Eq.~(\ref{eq:memory_original}) can be written as
\begin{equation}
    \langle \tau_i\tau_{i+1}\rangle = \sum_{\mu,\nu\in \{0,1\}} t_{\mu\nu} \tau^{(\mu)} \tau^{(\nu)},
\end{equation}
where
\begin{equation}
    \tau^{(0)} \equiv \frac {\int_0^{\Delta t}\tau P(\tau)d\tau} {\int_0^{\Delta t}P(\tau)d\tau},\ 
    \tau^{(1)} \equiv \frac {\int_{\Delta t}^\infty \tau P(\tau)d\tau} {\int_{\Delta t}^\infty P(\tau)d\tau}.
    \label{eq:tau01_mean}
\end{equation}
Here we have assumed that the information on the correlation between $\tau_i$ and $\tau_{i+1}$ is carried only by $t_{\mu\nu}$, while such consecutive IETs are independent of each other under the condition that $\tau_i\in \mathcal{T}_\mu$ and $\tau_{i+1}\in \mathcal{T}_\nu$. This assumption of conditional independence is based on the fact that the correlation between $\tau_i$ and $\tau_{i+1}$ with $\tau_i\in \mathcal{T}_\mu$ and $\tau_{i+1}\in \mathcal{T}_\nu$ is no longer relevant to the burst size statistics, because the bursty trains are determined depending only on whether each IET is larger than $\Delta t$ or not. Then $M$ in Eq.~(\ref{eq:memory_original}) reads in the asymptotic limit with $n\gg 1$
\begin{equation}
    \label{eq:memory_approx}
    M\simeq \frac{ \sum_{\mu,\nu\in \{0,1\}} t_{\mu\nu} \tau^{(\mu)} \tau^{(\nu)} - \langle \tau\rangle^2}{\sigma^2}.
\end{equation}
Here we have approximated as $\langle \tau\rangle_1 \simeq \langle \tau\rangle_2 \simeq \langle \tau\rangle$ and $\sigma_1 \simeq \sigma_2 \simeq \sigma$, with $\langle\tau\rangle$ and $\sigma$ denoting the average and standard deviation of IETs, respectively. Note that $\tau^{(0)}$ and $\tau^{(1)}$ are related as follows:
\begin{equation}
    \label{eq:tau01_relation}
    \left(1-\frac{1}{\langle b\rangle}\right) \tau^{(0)} +\frac{1}{\langle b\rangle} \tau^{(1)} \simeq \langle \tau\rangle.
\end{equation}

For deriving $M$ in Eq.~(\ref{eq:memory_approx}), $t_{\mu\nu}$s need to be calculated. Since each pair of IETs in $\mathcal{T}_{11}$ implies a bursty train of size $1$, the average size of $\mathcal{T}_{11}$ is $mQ_{\Delta t}(1)$, with $m$ denoting the number of bursty trains detected using $\Delta t$. Thus, the average fraction of IET pairs in $\mathcal{T}_{11}$ becomes
\begin{equation}
    \label{eq:t11_single}
    t_{11}\equiv \frac{\langle |\mathcal{T}_{11}|\rangle}{n-1} \simeq \frac{Q_{\Delta t}(1)}{\langle b\rangle},
\end{equation}
where Eq.~(\ref{eq:mn_event}) has been used. The pair of IETs in $\mathcal{T}_{10}$ ($\mathcal{T}_{01}$) is found whenever a bursty train of size larger than $1$ begins (ends). Hence, the average fraction of $\mathcal{T}_{10}$, equivalent to that of $\mathcal{T}_{01}$, must be 
\begin{equation}
    \label{eq:t10_single}
    t_{10} \equiv \frac{\langle |\mathcal{T}_{10}|\rangle}{n-1} \simeq \frac{1}{\langle b\rangle}\sum_{b=2}^\infty Q_{\Delta t}(b) = \frac{1-Q_{\Delta t}(1)}{\langle b\rangle},
\end{equation}
which is the same as $t_{01} \equiv \langle |\mathcal{T}_{01}|\rangle/(n-1)$. Finally, for each bursty train of size larger than $2$, we find $b-2$ pairs of IETs belonging to $\mathcal{T}_{00}$, indicating that the average fraction of $\mathcal{T}_{00}$ is
\begin{equation}
    \label{eq:t00_single}
    t_{00} \equiv \frac{\langle |\mathcal{T}_{00}|\rangle}{n-1} \simeq \frac{1}{\langle b\rangle}\sum_{b=3}^\infty (b-2)Q_{\Delta t}(b) = \frac{\langle b\rangle -2 +Q_{\Delta t}(1)}{\langle b\rangle}.
\end{equation}
Note that $t_{00} + t_{01} + t_{10} + t_{11}\simeq 1$. Then by using Eqs.~(\ref{eq:tau01_mean}) and~(\ref{eq:tau01_relation}) one obtains
\begin{equation}
    \sum_{\mu,\nu\in \{0,1\}} t_{\mu\nu}\tau^{(\mu)} \tau^{(\nu)} = [\langle b\rangle Q_{\Delta t}(1)-1] ( \langle \tau\rangle - \tau^{(0)})^2 +\langle \tau\rangle^2, 
\end{equation}
finally leading to
\begin{equation}
    \label{eq:memory_result}
    M \simeq \frac{[\langle b\rangle Q_{\Delta t}(1)-1] ( \langle \tau\rangle - \tau^{(0)})^2}{\sigma^2}.
\end{equation}
This solution has been derived for arbitrary forms of $P(\tau)$ and $Q_{\Delta t}(b)$.

\begin{figure}[!t]
    \includegraphics[width=\columnwidth]{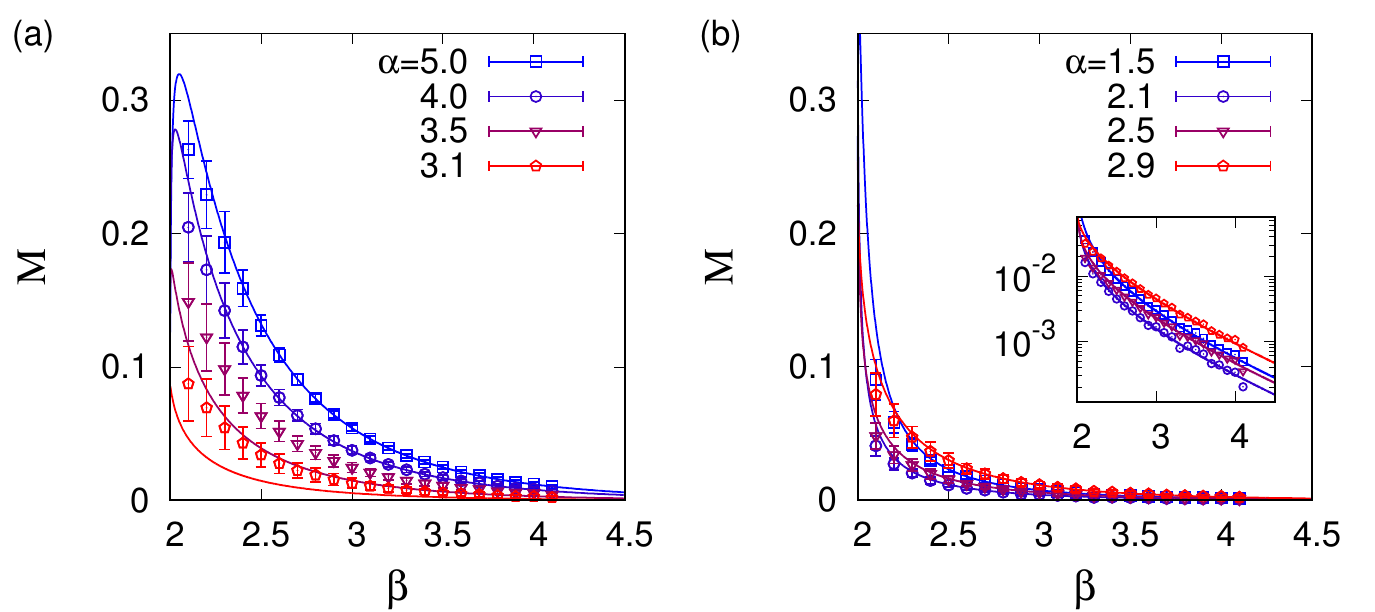}
    \caption{The analytical solution of $M$ in Eq.~(\ref{eq:memory_result}) as a function of $\beta$ in Eq.~(\ref{eq:Qb0}) for several values of $\alpha$ in Eq.~(\ref{eq:Ptau_cutoff}) (solid lines), compared with corresponding numerical results (symbols with error bars). In panel (a) we use the pure power-law distribution of $P(\tau)$ in Eq.~(\ref{eq:Ptau_cutoff}), with infinite exponential cutoff, i.e., $\tau_c\to\infty$, while the general form of $P(\tau)$ with $\tau_c=10^3\tau_{\rm min}$ is used in panel (b). The inset shows the same result as in panel (b), but in a semi-log scale. Each point and its standard deviation are obtained from $50$ event sequences of size $n=5\times 10^5$. Reprinted figure with permission from Ref.~\cite{Jo2018Limits} Copyright (2018) by the American Physical Society.}
    \label{fig:single_timescale}
\end{figure}

We investigate the dependence of $M$ on $Q_{\Delta t}(b)$, while keeping the same $P(\tau)$. As for the burst size distribution, we consider a power-law distribution as follows:
\begin{equation}
    \label{eq:Qb_powerlaw}
    Q_{\Delta t}(b) = \zeta(\beta)^{-1} b^{-\beta}\ \textrm{for}\ b=1,2,\cdots.
\end{equation}
We assume that $\beta>2$ for the existence of $\langle b\rangle$, i.e., $\langle b\rangle = \zeta(\beta-1)/\zeta(\beta)$. As for the IET distribution, a power-law distribution with an exponential cutoff is considered:
\begin{equation}
    \label{eq:Ptau_cutoff}
    P(\tau)= \frac{\tau_c^{\alpha-1}}{\Gamma(1-\alpha,\tau_{\rm min}/\tau_c)} \tau^{-\alpha} e^{-\tau/\tau_c} \theta(\tau-\tau_{\rm min}),
\end{equation}
where $\tau_{\rm min}$ and $\tau_c$ denote the lower bound and the exponential cutoff of $\tau$, respectively. Here $\Gamma(\cdot,\cdot)$ denotes the upper incomplete Gamma function. Figure~\ref{fig:single_timescale} shows how $M$ varies according to the power-law exponent $\beta$ for a given $\alpha$ for both cases with diverging and finite $\tau_c$, respectively. For the numerical simulations, the event sequences were generated using the implementation method of the BGB mechanism in Subsection~\ref{subsec:structure}, but using Eq.~(\ref{eq:Ptau_cutoff}). We confirm the tendency that the larger positive value of $M$ is associated with the smaller value of $\beta$, i.e., the heavier tail. This tendency can be understood by the intuition that the smaller $\beta$ implies the stronger correlations between IETs, possibly leading to the larger $M$. We also find that $M\approx 0$ for $\beta\approx 4$, whether $\tau_c$ is finite or infinite. This implies that the apparently conflicting observations in human activities are indeed compatible. Hence, we raise an important question regarding the effectiveness or limits of $M$ in measuring higher-order correlations between IETs. Although the definition of $M$ is straightforward and intuitive, it may not properly characterize the complex correlation structure between IETs in some cases.

\section{Effects of correlations between IETs on dynamical processes}\label{sec:spreading}

The dynamical processes, such as spreading, diffusion, and cascades, taking place in a temporal network of individuals are known to be strongly affected by bursty interaction patterns between individuals~\cite{Vazquez2007Impact, Karsai2011Small, Miritello2011Dynamical, Iribarren2009Impact, Rocha2011Simulated, Rocha2013Bursts, Takaguchi2013Bursty, Masuda2013Predicting, Jo2014Analytically, Perotti2014Temporal, Delvenne2015Diffusion,  Pastor-Satorras2015Epidemic, Artime2017Dynamics, Hiraoka2018Correlated}: In particular, spreading processes in temporal networks have been extensively studied. An important question is what features of temporal networks are most relevant to predict the speed of propagation, e.g., of disease or information. One of the crucial and widely studied features is the heterogeneities of IETs in the temporal interaction patterns. It was shown that the bursty interaction patterns can slow down the early-stage spreading by comparing the simulated spreading behaviors in some empirical networks and in their randomized versions~\cite{Vazquez2007Impact, Karsai2011Small, Perotti2014Temporal}. The opposite tendency was also reported using another empirical network or model networks~\cite{Rocha2011Simulated, Rocha2013Bursts, Jo2014Analytically}.

In contrast to the effects of heterogeneous IETs on the spreading, yet little is known about the effects of correlations between IETs on the spreading, except for few recent works~\cite{Artime2017Dynamics, Masuda2018Gillespie}. This could be partly because the contagion dynamics studied in many previous works, e.g., susceptible-infected (SI) dynamics~\cite{Pastor-Satorras2015Epidemic}, has focused on an immediate infection upon the first contact between susceptible and infected nodes, hence without the need to consider correlated IETs. In another work~\cite{Gueuning2015Imperfect}, probabilistic contagion dynamics, which naturally involves multiple consecutive IETs, was studied by assuming heterogeneous but uncorrelated IETs. Therefore, the effects of heterogeneous and correlated IETs on the spreading need to be systematically studied for better understanding the dynamical processes in complex systems.

\begin{figure}[tb]
    \includegraphics[width=0.9\columnwidth]{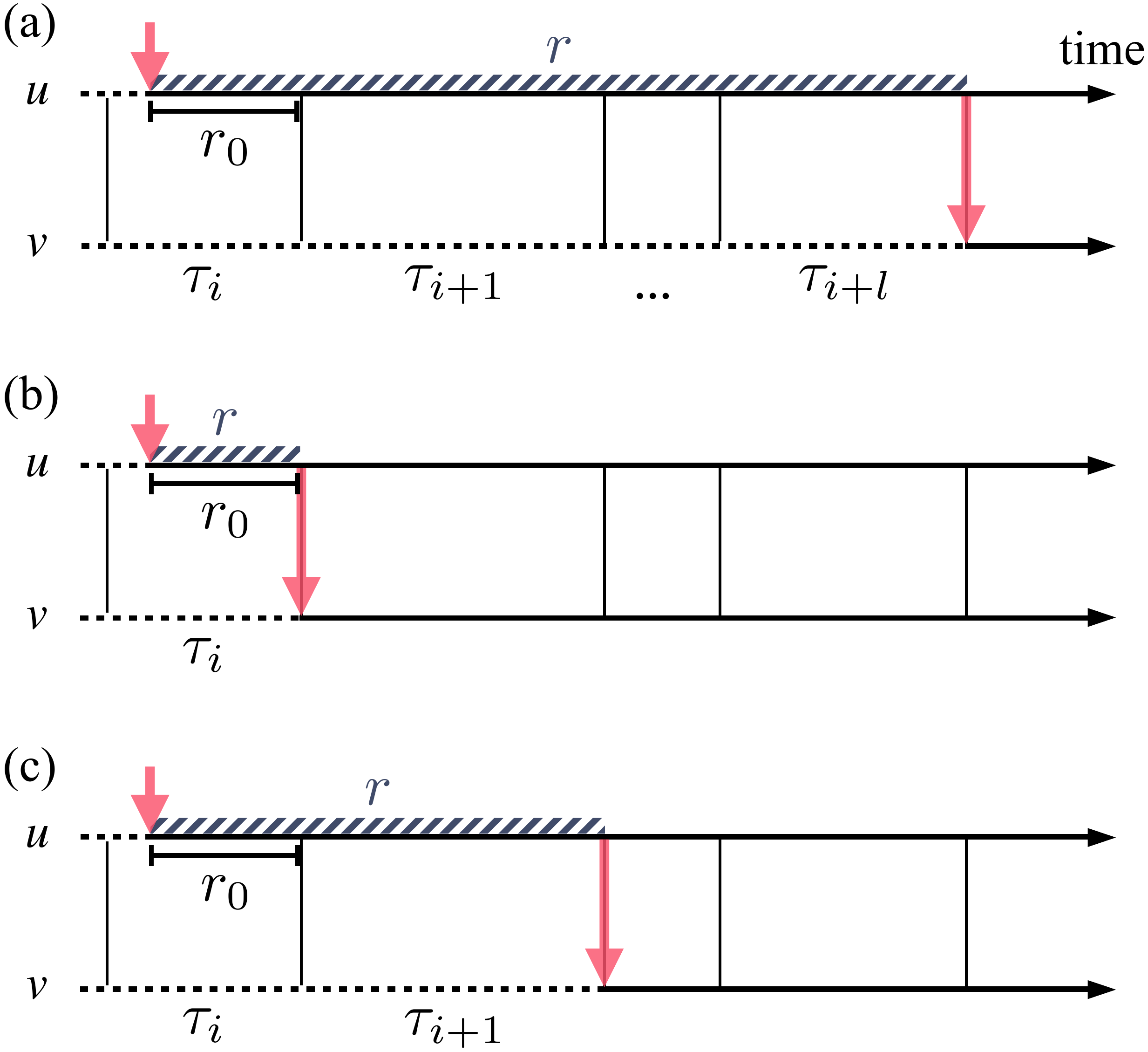}
    \caption{Schematic diagrams for (a) the probabilistic susceptible-infected (SI) dynamics, (b) the one-step deterministic SI dynamics, and (c) the two-step deterministic SI dynamics. For each node, the susceptible or intermediate state is represented by a dashed horizontal line, while the infected state is by a solid horizontal line. In each panel, a node $u$ gets infected in the time denoted by an upper vertical arrow, then it tries to infect its susceptible neighbor $v$ whenever they make contact (vertical lines). The successful infection of $v$ by $u$ is marked by a lower vertical arrow. The time interval between the infection of $u$ and that of $v$ (striped band) defines the transmission time $r$. For the definitions of $r_0$ and $\tau$s, see the text. Figure in Ref.~\cite{Hiraoka2018Correlated} by Takayuki Hiraoka and Hang-Hyun Jo is licensed under CC BY 4.0.}
    \label{fig:spreading}
\end{figure}

To study the spreading dynamics, one can consider one of the extensively studied epidemic processes, i.e., susceptible-infected (SI) dynamics~\cite{Pastor-Satorras2015Epidemic}: A state of each node in a network is either susceptible or infected, and an infected node can infect a susceptible node by the contact with it. Here we assume that the contact is instantaneous. One can study a probabilistic SI dynamics, in which an infected node can infect a susceptible node with probability $\eta$ ($0 < \eta < 1$) per contact, as depicted in Fig.~\ref{fig:spreading}(a). Due to the stochastic nature of infection, multiple IETs can be involved in the contagion, hence the correlations between IETs in the contact patterns can influence the spreading behavior. The case with $\eta=1$ corresponds to the deterministic version of SI dynamics: A susceptible node is immediately infected after its first contact with an infected node, see Fig.~\ref{fig:spreading}(b). Finally, for studying the effect of correlations between IETs on the spreading in a simpler setup, we introduce two-step deterministic SI (``2DSI'' in short) dynamics~\cite{Hiraoka2018Correlated} as a variation of generalized epidemic processes~\cite{Janssen2004Generalized, Dodds2004Universal, Bizhani2012Discontinuous, Chung2014Generalized}, see Fig.~\ref{fig:spreading}(c). Here a susceptible node first changes its state to an intermediate state upon its first contact with an infected node; it then becomes infected after the second contact with the same or another infected node. Below we only introduce the results for 2DSI dynamics from Ref.~\cite{Hiraoka2018Correlated}.

\begin{figure*}[!t]
    \includegraphics[width=0.92\linewidth]{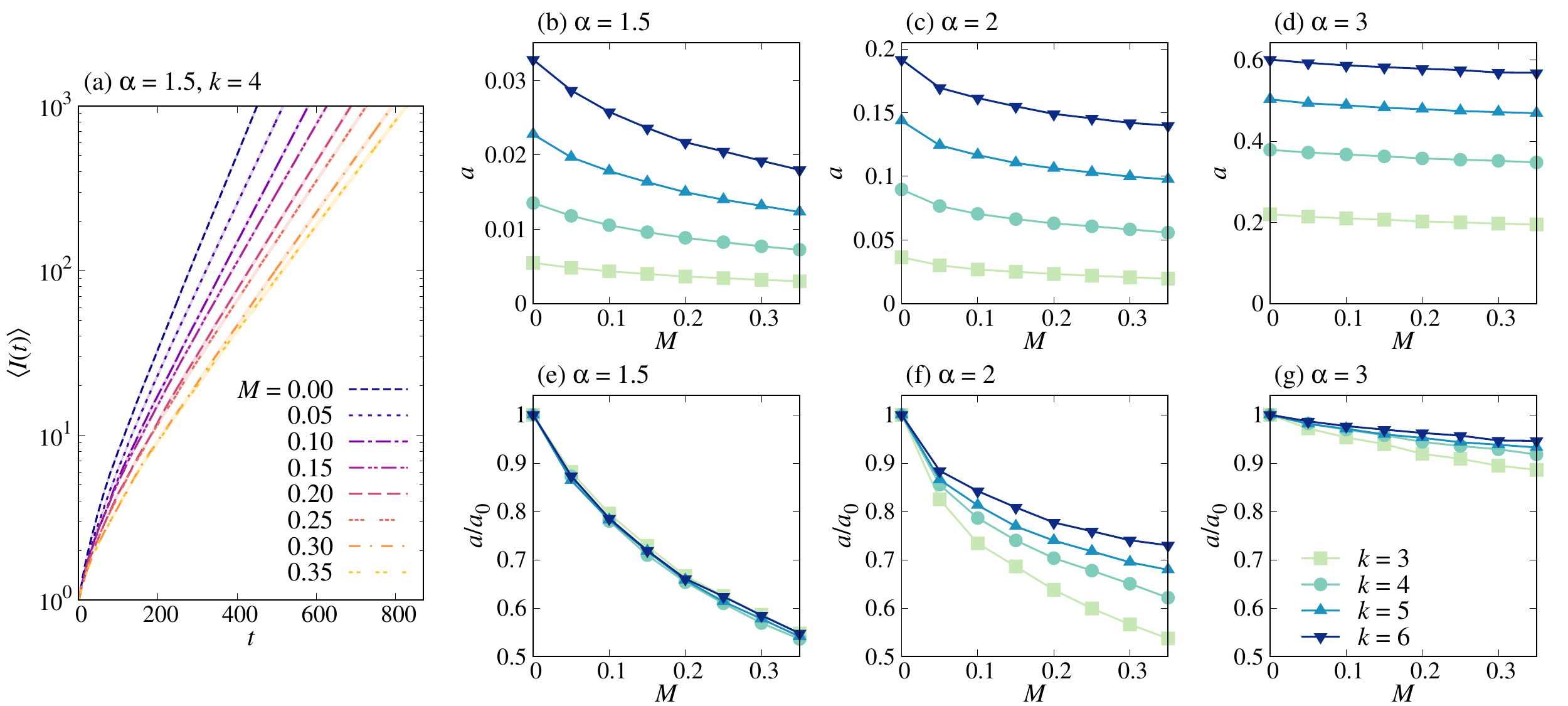}
    \caption{Two-step deterministic SI dynamics in Bethe lattices: (a) Average numbers of infected nodes as a function of time, $\langle I(t)\rangle$, in Bethe lattices with $k=4$ for the same IET distribution with power-law exponent $\alpha = 1.5$ in Eq.~(\ref{eq:Ptau_cutoff}), but with various values of memory coefficient $M$. For each value of $M$, the average (dashed curve) and its standard error (shaded area) were obtained from $10^3$ runs with different initial conditions. (b--g) Estimated exponential growth rates $a$, defined in Eq.~(\ref{eq:spreading}) (top panels) and their relative growth rates $a/a_0$ with $a_0\equiv a(M=0)$ (bottom panels) are plotted for various values of $k$, $\alpha$, and $M$. The lines are guides to the eye. Figure in Ref.~\cite{Hiraoka2018Correlated} by Takayuki Hiraoka and Hang-Hyun Jo is licensed under CC BY 4.0.}
    \label{fig:2dsi}
\end{figure*} 

For modeling the interaction structure in a population, we focus on Bethe lattices as networks of infinite size, where each node has $k$ neighbors. As for the temporal contact patterns, we assume that the contacts between a pair of nodes or on a link connecting these nodes are instantaneous and undirected. Moreover, the contact pattern on each link is assumed to be independent of the states of two end nodes as well as of contact patterns on other links. The contact pattern on each link is modeled by a statistically identical event sequence with heterogeneous and correlated IETs. For this, the shape of IET distribution $P(\tau)$ and the value of memory coefficient $M$ are given as inputs of the model. As for the IET distribution, we adopt $P(\tau)$ in Eq.~(\ref{eq:Ptau_cutoff}). We fix $\tau_{\min} = 1$ without loss of generality and set $\tau_c = 10^3$ in our work. Based on the empirical findings for $\alpha$~\cite{Karsai2018Bursty}, we consider the case with $1.5\leq \alpha\leq 3$. Secondly, only the positive memory coefficient $M$ is considered, precisely, $0\leq M< 0.4$, based on the empirical observations~\cite{Goh2008Burstiness, Wang2015Temporal, Guo2017Bounds, Bottcher2017Temporal}. 

Precisely, for each link, we draw $n$ random values from $P(\tau)$ to make an IET sequence $\mathcal{T}= \{\tau_1,\cdots, \tau_n\}$, for sufficiently large $n$. Using Eq.~(\ref{eq:memory_original}), we measure the memory coefficient from $\mathcal{T}$, denoted by $\tilde M$. Two IETs are randomly chosen in $\mathcal{T}$ and swapped only when this swapping makes $\tilde M$ closer to $M$, i.e., the target value. By repeating the swapping, we obtain the IET sequence whose $\tilde M$ is close enough to $M$, and from this IET sequence we get the sequence of contact timings for each link\footnote{Another algorithm for generating bursty time series using the copula has recently been suggested~\cite{Jo2019Copulabased}.}. Then the temporal network can be fully described by a set of contact timings for all links. Each simulation begins with one node infected at random in time, which we set as $t=0$, while all other nodes are susceptible at this moment. For each simulation, we measure the number of infected nodes as a function of time, $I(t)$. The average number of infected nodes $\langle I(t) \rangle$ is found to exponentially increase with time, e.g., as shown in Fig.~\ref{fig:2dsi}(a):
\begin{equation}
    \langle I(t) \rangle \sim e^{at}, 
    \label{eq:spreading}
\end{equation}
where $a=a(k,\alpha,M)$ denotes the exponential growth rate, known as the Malthusian parameter~\cite{Kimmel2002Branching}. $a$ turns out to be a decreasing function of $M$, indicating the slowdown of spreading due to the positive correlation between IETs, see Fig.~\ref{fig:2dsi}(b--d). The slowdown can be more clearly presented in terms of the relative growth rate $a/a_0$ with $a_0 \equiv a(M=0)$ for all cases of $k$ and $\alpha$, as shown in Fig.~\ref{fig:2dsi}(e--g). We summarize the main observations from the numerical simulations as follows:
\begin{enumerate}
    \item{$a$ decreases with $M$.}
    \item{$a$ increases with $\alpha$.}
    \item{$a$ increases with $k$.}
    \item{The deviation of $a/a_0$ from $1$ tends to be larger for smaller $\alpha$.}
\end{enumerate}

For understanding these observations, we provide an analytical solution for the transmission time in a single link setup. Let us consider a link connecting nodes $u$ and $v$, see Fig.~\ref{fig:spreading}. If $u$ gets infected from its neighbor other than $v$ in time $t_u$, and later it infects $v$ in time $t_v$, the time interval between $t_u$ and $t_v$ defines the transmission time $r\equiv t_v-t_u$. Here we assume that $v$ is not affected by any other neighbors than $u$, for the sake of simplicity. In order for the infected $u$ to infect the susceptible $v$, $u$ must wait at least for the next contact with $v$. This waiting or residual time is denoted by $r_0$. For the 2DSI dynamics, the transmission process involves two consecutive IETs. If the infection of $u$ occurs during the IET of $\tau_i$, then the transmission time is written as 
\begin{equation}
    \label{eq:r_2D}
    r=r_0+\tau_{i+1},
\end{equation}
with $\tau_{i+1}$ denoting the IET following $\tau_i$. Information on the correlations between $\tau_i$ and $\tau_{i+1}$ is carried by the joint distribution $P(\tau_i,\tau_{i+1})$ or the conditional distribution $P(\tau_{i+1}|\tau_i)$. Using $P(\tau_{i+1}|\tau_i)$ with $\tau_{i+1}=r-r_0$, the transmission time distribution is written as
\begin{equation}
    R(r) = \frac{1}{\langle\tau\rangle} \int_0^r dr_0 \int_{r_0}^\infty d\tau_i P(\tau_i) P(r-r_0| \tau_i),
    \label{eq:Qr_2D}
\end{equation}
where it is obvious from Eq.~(\ref{eq:r_2D}) that $\tau_i\geq r_0$ and $0\leq r_0\leq r$. The average transmission time is calculated as 
\begin{equation}
    \langle r\rangle \equiv \int_0^\infty dr rR(r)= \frac{1}{2}\left(\langle\tau\rangle + \frac{\sigma^2}{\langle\tau\rangle}\right) +\frac{1}{\langle\tau\rangle} \langle \tau_i\tau_{i+1}\rangle,
\end{equation}
where
\begin{equation}
    \langle \tau_i\tau_{i+1}\rangle\equiv \int_0^\infty d\tau_i \int_0^\infty d\tau_{i+1} \tau_i\tau_{i+1} P(\tau_i,\tau_{i+1}).
    \label{eq:tautau1}
\end{equation}
In order to relate this result to the memory coefficient in Eq.~(\ref{eq:memory_original}), we define a parameter as
\begin{equation}
    M \equiv \frac{\langle \tau_i \tau_{i+1} \rangle - \langle\tau\rangle^2}{\sigma^2}
    \label{eq:M_approx}
\end{equation}
to finally obtain the analytical result of the average transmission time: 
\begin{equation}
    \label{eq:avg_r_2D}
    \langle r\rangle = \frac{3}{2} \langle\tau\rangle + \left(\frac{1}{2} + M\right) \frac{\sigma^2}{\langle\tau\rangle}.
\end{equation}

We remark that our result in Eq.~(\ref{eq:avg_r_2D}) is valid for arbitrary functional forms of IET distributions as long as their mean and variance are finite. $M$ is coupled with $\sigma^2/\langle\tau\rangle$, implying that the impact of correlations between IETs becomes larger with broader IET distributions. More importantly, we find that a stronger positive correlation between consecutive IETs leads to a larger average transmission time. This can be understood in terms of the role of the variance of IETs in the average transmission time. That is, the variance of the sum of two consecutive IETs is amplified by the positive correlation between those IETs. Based on the result of the single-link analysis, we can undersand the numerical results in Fig.~\ref{fig:2dsi}: The decreasing $a$ with $M$ is expected from Eq.~(\ref{eq:avg_r_2D}), so is the increasing $a$ with $\alpha$ as both $\langle\tau\rangle$ and $\sigma^2/\langle\tau\rangle$ decrease with $\alpha$. The observation that the deviation of $a/a_0$ from $1$ tends to be larger for smaller $\alpha$ implies that the effect of $M$ becomes larger for smaller $\alpha$, which can be roughly understood by a larger value of $\sigma^2/\langle\tau\rangle$ coupled to $M$ in Eq.~(\ref{eq:avg_r_2D}). Finally, the increasing $a$ with the degree $k$ is trivial, while the analytical approach to this dependency is not trivial, calling for more rigorous investigation.

\section{Discussion}\label{sec:discuss}

In this Chapter, we have introduced various measures and characterizations for bursty time series analysis and showed how they can be related to each other. Yet more rigorous studies need to be done for understanding such relation comprehensively. In the context of temporal networks, the superposition of event sequences of links incident to a node can result in the event sequence of the node. Then bursty behaviors of a node can be understood in terms of those of links incident to the node. For analyzing the relation between bursty behaviors in nodes and links, one can adopt the notion of contextual bursts by which the scaling behaviors of IET distributions of nodes and links can be systematically understood~\cite{Jo2013Contextual}. Researchers can also study how the correlations between IETs in one node or link are related to those in other nodes or links, how such inter-correlations can be properly characterized, and how they can affect the dynamical processes taking place in temporal networks. 

\begin{acknowledgments}
  The authors acknowledge financial support by Basic Science Research Program through the National Research Foundation of Korea (NRF) grant funded by the Ministry of Education (NRF-2018R1D1A1A09081919).
\end{acknowledgments}

\bibliographystyle{apsrev4-1}

\begin{thebibliography}{64}%
    \makeatletter
    \providecommand \@ifxundefined [1]{%
     \@ifx{#1\undefined}
    }%
    \providecommand \@ifnum [1]{%
     \ifnum #1\expandafter \@firstoftwo
     \else \expandafter \@secondoftwo
     \fi
    }%
    \providecommand \@ifx [1]{%
     \ifx #1\expandafter \@firstoftwo
     \else \expandafter \@secondoftwo
     \fi
    }%
    \providecommand \natexlab [1]{#1}%
    \providecommand \enquote  [1]{``#1''}%
    \providecommand \bibnamefont  [1]{#1}%
    \providecommand \bibfnamefont [1]{#1}%
    \providecommand \citenamefont [1]{#1}%
    \providecommand \href@noop [0]{\@secondoftwo}%
    \providecommand \href [0]{\begingroup \@sanitize@url \@href}%
    \providecommand \@href[1]{\@@startlink{#1}\@@href}%
    \providecommand \@@href[1]{\endgroup#1\@@endlink}%
    \providecommand \@sanitize@url [0]{\catcode `\\12\catcode `\$12\catcode
      `\&12\catcode `\#12\catcode `\^12\catcode `\_12\catcode `\%12\relax}%
    \providecommand \@@startlink[1]{}%
    \providecommand \@@endlink[0]{}%
    \providecommand \url  [0]{\begingroup\@sanitize@url \@url }%
    \providecommand \@url [1]{\endgroup\@href {#1}{\urlprefix }}%
    \providecommand \urlprefix  [0]{URL }%
    \providecommand \Eprint [0]{\href }%
    \providecommand \doibase [0]{https://doi.org/}%
    \providecommand \selectlanguage [0]{\@gobble}%
    \providecommand \bibinfo  [0]{\@secondoftwo}%
    \providecommand \bibfield  [0]{\@secondoftwo}%
    \providecommand \translation [1]{[#1]}%
    \providecommand \BibitemOpen [0]{}%
    \providecommand \bibitemStop [0]{}%
    \providecommand \bibitemNoStop [0]{.\EOS\space}%
    \providecommand \EOS [0]{\spacefactor3000\relax}%
    \providecommand \BibitemShut  [1]{\csname bibitem#1\endcsname}%
    \let\auto@bib@innerbib\@empty
    \bibitem [{\citenamefont {Albert}\ and\ \citenamefont
      {Barab{\'a}si}(2002)}]{Albert2002Statistical}%
      \BibitemOpen
      \bibfield  {author} {\bibinfo {author} {\bibfnamefont {R.}~\bibnamefont
      {Albert}}\ and\ \bibinfo {author} {\bibfnamefont {A.-L.}\ \bibnamefont
      {Barab{\'a}si}},\ }\href {https://doi.org/10.1103/revmodphys.74.47}
      {\bibfield  {journal} {\bibinfo  {journal} {Review of Modern Physics}\
      }\textbf {\bibinfo {volume} {74}},\ \bibinfo {pages} {47} (\bibinfo {year}
      {2002})}\BibitemShut {NoStop}%
    \bibitem [{\citenamefont {Newman}(2010)}]{Newman2010Networks}%
      \BibitemOpen
      \bibfield  {author} {\bibinfo {author} {\bibfnamefont {M.~E.~J.}\
      \bibnamefont {Newman}},\ }\href {http://www.worldcat.org/isbn/0199206651}
      {\emph {\bibinfo {title} {Networks: {An} {Introduction}}}},\ \bibinfo
      {edition} {1st}\ ed.\ (\bibinfo  {publisher} {Oxford University Press},\
      \bibinfo {year} {2010})\BibitemShut {NoStop}%
    \bibitem [{\citenamefont {Holme}\ and\ \citenamefont
      {Saram{\"a}ki}(2012)}]{Holme2012Temporal}%
      \BibitemOpen
      \bibfield  {author} {\bibinfo {author} {\bibfnamefont {P.}~\bibnamefont
      {Holme}}\ and\ \bibinfo {author} {\bibfnamefont {J.}~\bibnamefont
      {Saram{\"a}ki}},\ }\href {https://doi.org/10.1016/j.physrep.2012.03.001}
      {\bibfield  {journal} {\bibinfo  {journal} {Physics Reports}\ }\textbf
      {\bibinfo {volume} {519}},\ \bibinfo {pages} {97} (\bibinfo {year}
      {2012})}\BibitemShut {NoStop}%
    \bibitem [{\citenamefont {Masuda}\ and\ \citenamefont
      {Lambiotte}(2016)}]{Masuda2016Guide}%
      \BibitemOpen
      \bibfield  {author} {\bibinfo {author} {\bibfnamefont {N.}~\bibnamefont
      {Masuda}}\ and\ \bibinfo {author} {\bibfnamefont {R.}~\bibnamefont
      {Lambiotte}},\ }\href@noop {} {\emph {\bibinfo {title} {A guide to temporal
      networks}}},\ Series on complexity science\ (\bibinfo  {publisher} {World
      Scientific},\ \bibinfo {address} {New Jersey},\ \bibinfo {year}
      {2016})\BibitemShut {NoStop}%
    \bibitem [{\citenamefont {Gauvin}\ \emph {et~al.}(2018)\citenamefont {Gauvin},
      \citenamefont {G{\'e}nois}, \citenamefont {Karsai}, \citenamefont
      {Kivel{\"a}}, \citenamefont {Takaguchi}, \citenamefont {Valdano},\ and\
      \citenamefont {Vestergaard}}]{Gauvin2018Randomized}%
      \BibitemOpen
      \bibfield  {author} {\bibinfo {author} {\bibfnamefont {L.}~\bibnamefont
      {Gauvin}}, \bibinfo {author} {\bibfnamefont {M.}~\bibnamefont {G{\'e}nois}},
      \bibinfo {author} {\bibfnamefont {M.}~\bibnamefont {Karsai}}, \bibinfo
      {author} {\bibfnamefont {M.}~\bibnamefont {Kivel{\"a}}}, \bibinfo {author}
      {\bibfnamefont {T.}~\bibnamefont {Takaguchi}}, \bibinfo {author}
      {\bibfnamefont {E.}~\bibnamefont {Valdano}},\ and\ \bibinfo {author}
      {\bibfnamefont {C.~L.}\ \bibnamefont {Vestergaard}}} (\bibinfo {year}
      {2018}),\ \bibinfo {note} {arXiv:1806.04032}\BibitemShut {NoStop}%
    \bibitem [{\citenamefont {Barab{\'a}si}(2005)}]{Barabasi2005Origin}%
      \BibitemOpen
      \bibfield  {author} {\bibinfo {author} {\bibfnamefont {A.-L.}\ \bibnamefont
      {Barab{\'a}si}},\ }\href {https://doi.org/10.1038/nature03459} {\bibfield
      {journal} {\bibinfo  {journal} {Nature}\ }\textbf {\bibinfo {volume} {435}},\
      \bibinfo {pages} {207} (\bibinfo {year} {2005})}\BibitemShut {NoStop}%
    \bibitem [{\citenamefont {Karsai}\ \emph
      {et~al.}(2012{\natexlab{a}})\citenamefont {Karsai}, \citenamefont {Kaski},
      \citenamefont {Barab{\'a}si},\ and\ \citenamefont
      {Kert{\'e}sz}}]{Karsai2012Universal}%
      \BibitemOpen
      \bibfield  {author} {\bibinfo {author} {\bibfnamefont {M.}~\bibnamefont
      {Karsai}}, \bibinfo {author} {\bibfnamefont {K.}~\bibnamefont {Kaski}},
      \bibinfo {author} {\bibfnamefont {A.-L.}\ \bibnamefont {Barab{\'a}si}},\ and\
      \bibinfo {author} {\bibfnamefont {J.}~\bibnamefont {Kert{\'e}sz}},\ }\href
      {https://doi.org/10.1038/srep00397} {\bibfield  {journal} {\bibinfo
      {journal} {Scientific Reports}\ }\textbf {\bibinfo {volume} {2}},\ \bibinfo
      {pages} {397} (\bibinfo {year} {2012}{\natexlab{a}})}\BibitemShut {NoStop}%
    \bibitem [{\citenamefont {Karsai}\ \emph {et~al.}(2018)\citenamefont {Karsai},
      \citenamefont {Jo},\ and\ \citenamefont {Kaski}}]{Karsai2018Bursty}%
      \BibitemOpen
      \bibfield  {author} {\bibinfo {author} {\bibfnamefont {M.}~\bibnamefont
      {Karsai}}, \bibinfo {author} {\bibfnamefont {H.-H.}\ \bibnamefont {Jo}},\
      and\ \bibinfo {author} {\bibfnamefont {K.}~\bibnamefont {Kaski}},\ }\href
      {http://www.worldcat.org/isbn/3319685384} {\emph {\bibinfo {title} {Bursty
      {Human} {Dynamics}}}}\ (\bibinfo  {publisher} {Springer International
      Publishing},\ \bibinfo {address} {Cham},\ \bibinfo {year} {2018})\BibitemShut
      {NoStop}%
    \bibitem [{\citenamefont {Eckmann}\ \emph {et~al.}(2004)\citenamefont
      {Eckmann}, \citenamefont {Moses},\ and\ \citenamefont
      {Sergi}}]{Eckmann2004Entropy}%
      \BibitemOpen
      \bibfield  {author} {\bibinfo {author} {\bibfnamefont {J.-P.}\ \bibnamefont
      {Eckmann}}, \bibinfo {author} {\bibfnamefont {E.}~\bibnamefont {Moses}},\
      and\ \bibinfo {author} {\bibfnamefont {D.}~\bibnamefont {Sergi}},\ }\href
      {https://doi.org/10.1073/pnas.0405728101} {\bibfield  {journal} {\bibinfo
      {journal} {Proceedings of the National Academy of Sciences}\ }\textbf
      {\bibinfo {volume} {101}},\ \bibinfo {pages} {14333} (\bibinfo {year}
      {2004})}\BibitemShut {NoStop}%
    \bibitem [{\citenamefont {Malmgren}\ \emph {et~al.}(2009)\citenamefont
      {Malmgren}, \citenamefont {Stouffer}, \citenamefont {Campanharo},\ and\
      \citenamefont {Amaral}}]{Malmgren2009Universality}%
      \BibitemOpen
      \bibfield  {author} {\bibinfo {author} {\bibfnamefont {R.~D.}\ \bibnamefont
      {Malmgren}}, \bibinfo {author} {\bibfnamefont {D.~B.}\ \bibnamefont
      {Stouffer}}, \bibinfo {author} {\bibfnamefont {A.~S. L.~O.}\ \bibnamefont
      {Campanharo}},\ and\ \bibinfo {author} {\bibfnamefont {L.~A.}\ \bibnamefont
      {Amaral}},\ }\href {https://doi.org/10.1126/science.1174562} {\bibfield
      {journal} {\bibinfo  {journal} {Science}\ }\textbf {\bibinfo {volume}
      {325}},\ \bibinfo {pages} {1696} (\bibinfo {year} {2009})}\BibitemShut
      {NoStop}%
    \bibitem [{\citenamefont {Cattuto}\ \emph {et~al.}(2010)\citenamefont
      {Cattuto}, \citenamefont {Van~den Broeck}, \citenamefont {Barrat},
      \citenamefont {Colizza}, \citenamefont {Pinton},\ and\ \citenamefont
      {Vespignani}}]{Cattuto2010Dynamics}%
      \BibitemOpen
      \bibfield  {author} {\bibinfo {author} {\bibfnamefont {C.}~\bibnamefont
      {Cattuto}}, \bibinfo {author} {\bibfnamefont {W.}~\bibnamefont {Van~den
      Broeck}}, \bibinfo {author} {\bibfnamefont {A.}~\bibnamefont {Barrat}},
      \bibinfo {author} {\bibfnamefont {V.}~\bibnamefont {Colizza}}, \bibinfo
      {author} {\bibfnamefont {J.-F.}\ \bibnamefont {Pinton}},\ and\ \bibinfo
      {author} {\bibfnamefont {A.}~\bibnamefont {Vespignani}},\ }\href
      {https://doi.org/10.1371/journal.pone.0011596} {\bibfield  {journal}
      {\bibinfo  {journal} {PLoS ONE}\ }\textbf {\bibinfo {volume} {5}},\ \bibinfo
      {pages} {e11596} (\bibinfo {year} {2010})}\BibitemShut {NoStop}%
    \bibitem [{\citenamefont {Jo}\ \emph {et~al.}(2012)\citenamefont {Jo},
      \citenamefont {Karsai}, \citenamefont {Kert{\'e}sz},\ and\ \citenamefont
      {Kaski}}]{Jo2012Circadian}%
      \BibitemOpen
      \bibfield  {author} {\bibinfo {author} {\bibfnamefont {H.-H.}\ \bibnamefont
      {Jo}}, \bibinfo {author} {\bibfnamefont {M.}~\bibnamefont {Karsai}}, \bibinfo
      {author} {\bibfnamefont {J.}~\bibnamefont {Kert{\'e}sz}},\ and\ \bibinfo
      {author} {\bibfnamefont {K.}~\bibnamefont {Kaski}},\ }\href
      {https://doi.org/10.1088/1367-2630/14/1/013055} {\bibfield  {journal}
      {\bibinfo  {journal} {New Journal of Physics}\ }\textbf {\bibinfo {volume}
      {14}},\ \bibinfo {pages} {013055} (\bibinfo {year} {2012})}\BibitemShut
      {NoStop}%
    \bibitem [{\citenamefont {Rybski}\ \emph {et~al.}(2012)\citenamefont {Rybski},
      \citenamefont {Buldyrev}, \citenamefont {Havlin}, \citenamefont {Liljeros},\
      and\ \citenamefont {Makse}}]{Rybski2012Communication}%
      \BibitemOpen
      \bibfield  {author} {\bibinfo {author} {\bibfnamefont {D.}~\bibnamefont
      {Rybski}}, \bibinfo {author} {\bibfnamefont {S.~V.}\ \bibnamefont
      {Buldyrev}}, \bibinfo {author} {\bibfnamefont {S.}~\bibnamefont {Havlin}},
      \bibinfo {author} {\bibfnamefont {F.}~\bibnamefont {Liljeros}},\ and\
      \bibinfo {author} {\bibfnamefont {H.~A.}\ \bibnamefont {Makse}},\ }\href
      {https://doi.org/10.1038/srep00560} {\bibfield  {journal} {\bibinfo
      {journal} {Scientific Reports}\ }\textbf {\bibinfo {volume} {2}},\ \bibinfo
      {pages} {560} (\bibinfo {year} {2012})}\BibitemShut {NoStop}%
    \bibitem [{\citenamefont {Jiang}\ \emph {et~al.}(2013)\citenamefont {Jiang},
      \citenamefont {Xie}, \citenamefont {Li}, \citenamefont {Podobnik},
      \citenamefont {Zhou},\ and\ \citenamefont {Stanley}}]{Jiang2013Calling}%
      \BibitemOpen
      \bibfield  {author} {\bibinfo {author} {\bibfnamefont {Z.-Q.}\ \bibnamefont
      {Jiang}}, \bibinfo {author} {\bibfnamefont {W.-J.}\ \bibnamefont {Xie}},
      \bibinfo {author} {\bibfnamefont {M.-X.}\ \bibnamefont {Li}}, \bibinfo
      {author} {\bibfnamefont {B.}~\bibnamefont {Podobnik}}, \bibinfo {author}
      {\bibfnamefont {W.-X.}\ \bibnamefont {Zhou}},\ and\ \bibinfo {author}
      {\bibfnamefont {H.~E.}\ \bibnamefont {Stanley}},\ }\href
      {https://doi.org/10.1073/pnas.1220433110} {\bibfield  {journal} {\bibinfo
      {journal} {Proceedings of the National Academy of Sciences}\ }\textbf
      {\bibinfo {volume} {110}},\ \bibinfo {pages} {1600} (\bibinfo {year}
      {2013})}\BibitemShut {NoStop}%
    \bibitem [{\citenamefont {Stopczynski}\ \emph {et~al.}(2014)\citenamefont
      {Stopczynski}, \citenamefont {Sekara}, \citenamefont {Sapiezynski},
      \citenamefont {Cuttone}, \citenamefont {Madsen}, \citenamefont {Larsen},\
      and\ \citenamefont {Lehmann}}]{Stopczynski2014Measuring}%
      \BibitemOpen
      \bibfield  {author} {\bibinfo {author} {\bibfnamefont {A.}~\bibnamefont
      {Stopczynski}}, \bibinfo {author} {\bibfnamefont {V.}~\bibnamefont {Sekara}},
      \bibinfo {author} {\bibfnamefont {P.}~\bibnamefont {Sapiezynski}}, \bibinfo
      {author} {\bibfnamefont {A.}~\bibnamefont {Cuttone}}, \bibinfo {author}
      {\bibfnamefont {M.~M.}\ \bibnamefont {Madsen}}, \bibinfo {author}
      {\bibfnamefont {J.~E.}\ \bibnamefont {Larsen}},\ and\ \bibinfo {author}
      {\bibfnamefont {S.}~\bibnamefont {Lehmann}},\ }\href
      {https://doi.org/10.1371/journal.pone.0095978} {\bibfield  {journal}
      {\bibinfo  {journal} {PLoS ONE}\ }\textbf {\bibinfo {volume} {9}},\ \bibinfo
      {pages} {e95978} (\bibinfo {year} {2014})}\BibitemShut {NoStop}%
    \bibitem [{\citenamefont {Panzarasa}\ and\ \citenamefont
      {Bonaventura}(2015)}]{Panzarasa2015Emergence}%
      \BibitemOpen
      \bibfield  {author} {\bibinfo {author} {\bibfnamefont {P.}~\bibnamefont
      {Panzarasa}}\ and\ \bibinfo {author} {\bibfnamefont {M.}~\bibnamefont
      {Bonaventura}},\ }\href {https://doi.org/10.1103/physreve.92.062821}
      {\bibfield  {journal} {\bibinfo  {journal} {Physical Review E}\ }\textbf
      {\bibinfo {volume} {92}},\ \bibinfo {pages} {062821} (\bibinfo {year}
      {2015})}\BibitemShut {NoStop}%
    \bibitem [{\citenamefont {Goh}\ and\ \citenamefont
      {Barab{\'a}si}(2008)}]{Goh2008Burstiness}%
      \BibitemOpen
      \bibfield  {author} {\bibinfo {author} {\bibfnamefont {K.-I.}\ \bibnamefont
      {Goh}}\ and\ \bibinfo {author} {\bibfnamefont {A.-L.}\ \bibnamefont
      {Barab{\'a}si}},\ }\href {https://doi.org/10.1209/0295-5075/81/48002}
      {\bibfield  {journal} {\bibinfo  {journal} {EPL (Europhysics Letters)}\
      }\textbf {\bibinfo {volume} {81}},\ \bibinfo {pages} {48002} (\bibinfo {year}
      {2008})}\BibitemShut {NoStop}%
    \bibitem [{\citenamefont {Jo}(2017)}]{Jo2017Modeling}%
      \BibitemOpen
      \bibfield  {author} {\bibinfo {author} {\bibfnamefont {H.-H.}\ \bibnamefont
      {Jo}},\ }\href {https://doi.org/10.1103/physreve.96.062131} {\bibfield
      {journal} {\bibinfo  {journal} {Physical Review E}\ }\textbf {\bibinfo
      {volume} {96}},\ \bibinfo {pages} {062131} (\bibinfo {year}
      {2017})}\BibitemShut {NoStop}%
    \bibitem [{\citenamefont {Wheatland}\ \emph {et~al.}(1998)\citenamefont
      {Wheatland}, \citenamefont {Sturrock},\ and\ \citenamefont
      {McTiernan}}]{Wheatland1998WaitingTime}%
      \BibitemOpen
      \bibfield  {author} {\bibinfo {author} {\bibfnamefont {M.~S.}\ \bibnamefont
      {Wheatland}}, \bibinfo {author} {\bibfnamefont {P.~A.}\ \bibnamefont
      {Sturrock}},\ and\ \bibinfo {author} {\bibfnamefont {J.~M.}\ \bibnamefont
      {McTiernan}},\ }\href {https://doi.org/10.1086/306492} {\bibfield  {journal}
      {\bibinfo  {journal} {The Astrophysical Journal}\ }\textbf {\bibinfo {volume}
      {509}},\ \bibinfo {pages} {448} (\bibinfo {year} {1998})}\BibitemShut
      {NoStop}%
    \bibitem [{\citenamefont {Corral}(2004)}]{Corral2004LongTerm}%
      \BibitemOpen
      \bibfield  {author} {\bibinfo {author} {\bibfnamefont {{\'A}.}~\bibnamefont
      {Corral}},\ }\href {https://doi.org/10.1103/physrevlett.92.108501} {\bibfield
       {journal} {\bibinfo  {journal} {Physical Review Letters}\ }\textbf {\bibinfo
      {volume} {92}},\ \bibinfo {pages} {108501} (\bibinfo {year}
      {2004})}\BibitemShut {NoStop}%
    \bibitem [{\citenamefont {de~Arcangelis}\ \emph {et~al.}(2006)\citenamefont
      {de~Arcangelis}, \citenamefont {Godano}, \citenamefont {Lippiello},\ and\
      \citenamefont {Nicodemi}}]{deArcangelis2006Universality}%
      \BibitemOpen
      \bibfield  {author} {\bibinfo {author} {\bibfnamefont {L.}~\bibnamefont
      {de~Arcangelis}}, \bibinfo {author} {\bibfnamefont {C.}~\bibnamefont
      {Godano}}, \bibinfo {author} {\bibfnamefont {E.}~\bibnamefont {Lippiello}},\
      and\ \bibinfo {author} {\bibfnamefont {M.}~\bibnamefont {Nicodemi}},\ }\href
      {https://doi.org/10.1103/physrevlett.96.051102} {\bibfield  {journal}
      {\bibinfo  {journal} {Physical Review Letters}\ }\textbf {\bibinfo {volume}
      {96}},\ \bibinfo {pages} {051102} (\bibinfo {year} {2006})}\BibitemShut
      {NoStop}%
    \bibitem [{\citenamefont {Kemuriyama}\ \emph {et~al.}(2010)\citenamefont
      {Kemuriyama}, \citenamefont {Ohta}, \citenamefont {Sato}, \citenamefont
      {Maruyama}, \citenamefont {Tandai-Hiruma}, \citenamefont {Kato},\ and\
      \citenamefont {Nishida}}]{Kemuriyama2010Powerlaw}%
      \BibitemOpen
      \bibfield  {author} {\bibinfo {author} {\bibfnamefont {T.}~\bibnamefont
      {Kemuriyama}}, \bibinfo {author} {\bibfnamefont {H.}~\bibnamefont {Ohta}},
      \bibinfo {author} {\bibfnamefont {Y.}~\bibnamefont {Sato}}, \bibinfo {author}
      {\bibfnamefont {S.}~\bibnamefont {Maruyama}}, \bibinfo {author}
      {\bibfnamefont {M.}~\bibnamefont {Tandai-Hiruma}}, \bibinfo {author}
      {\bibfnamefont {K.}~\bibnamefont {Kato}},\ and\ \bibinfo {author}
      {\bibfnamefont {Y.}~\bibnamefont {Nishida}},\ }\href
      {https://doi.org/10.1016/j.biosystems.2010.06.002} {\bibfield  {journal}
      {\bibinfo  {journal} {BioSystems}\ }\textbf {\bibinfo {volume} {101}},\
      \bibinfo {pages} {144} (\bibinfo {year} {2010})}\BibitemShut {NoStop}%
    \bibitem [{\citenamefont {Sorribes}\ \emph {et~al.}(2011)\citenamefont
      {Sorribes}, \citenamefont {Armendariz}, \citenamefont {Lopez-Pigozzi},
      \citenamefont {Murga},\ and\ \citenamefont
      {de~Polavieja}}]{Sorribes2011Origin}%
      \BibitemOpen
      \bibfield  {author} {\bibinfo {author} {\bibfnamefont {A.}~\bibnamefont
      {Sorribes}}, \bibinfo {author} {\bibfnamefont {B.~G.}\ \bibnamefont
      {Armendariz}}, \bibinfo {author} {\bibfnamefont {D.}~\bibnamefont
      {Lopez-Pigozzi}}, \bibinfo {author} {\bibfnamefont {C.}~\bibnamefont
      {Murga}},\ and\ \bibinfo {author} {\bibfnamefont {G.~G.}\ \bibnamefont
      {de~Polavieja}},\ }\href {https://doi.org/10.1371/journal.pcbi.1002075}
      {\bibfield  {journal} {\bibinfo  {journal} {PLoS Computational Biology}\
      }\textbf {\bibinfo {volume} {7}},\ \bibinfo {pages} {e1002075} (\bibinfo
      {year} {2011})}\BibitemShut {NoStop}%
    \bibitem [{\citenamefont {Boyer}\ \emph {et~al.}(2012)\citenamefont {Boyer},
      \citenamefont {Crofoot},\ and\ \citenamefont {Walsh}}]{Boyer2012Nonrandom}%
      \BibitemOpen
      \bibfield  {author} {\bibinfo {author} {\bibfnamefont {D.}~\bibnamefont
      {Boyer}}, \bibinfo {author} {\bibfnamefont {M.~C.}\ \bibnamefont {Crofoot}},\
      and\ \bibinfo {author} {\bibfnamefont {P.~D.}\ \bibnamefont {Walsh}},\ }\href
      {https://doi.org/10.1098/rsif.2011.0582} {\bibfield  {journal} {\bibinfo
      {journal} {Journal of The Royal Society Interface}\ }\textbf {\bibinfo
      {volume} {9}},\ \bibinfo {pages} {842} (\bibinfo {year} {2012})}\BibitemShut
      {NoStop}%
    \bibitem [{\citenamefont {Mainardi}\ \emph {et~al.}(2007)\citenamefont
      {Mainardi}, \citenamefont {Gorenflo},\ and\ \citenamefont
      {Vivoli}}]{Mainardi2007Poisson}%
      \BibitemOpen
      \bibfield  {author} {\bibinfo {author} {\bibfnamefont {F.}~\bibnamefont
      {Mainardi}}, \bibinfo {author} {\bibfnamefont {R.}~\bibnamefont {Gorenflo}},\
      and\ \bibinfo {author} {\bibfnamefont {A.}~\bibnamefont {Vivoli}},\ }\href
      {https://doi.org/10.1016/j.cam.2006.04.060} {\bibfield  {journal} {\bibinfo
      {journal} {Journal of Computational and Applied Mathematics}\ }\textbf
      {\bibinfo {volume} {205}},\ \bibinfo {pages} {725} (\bibinfo {year}
      {2007})}\BibitemShut {NoStop}%
    \bibitem [{\citenamefont {Lowen}\ and\ \citenamefont
      {Teich}(1993)}]{Lowen1993Fractal}%
      \BibitemOpen
      \bibfield  {author} {\bibinfo {author} {\bibfnamefont {S.~B.}\ \bibnamefont
      {Lowen}}\ and\ \bibinfo {author} {\bibfnamefont {M.~C.}\ \bibnamefont
      {Teich}},\ }\href {https://doi.org/10.1103/physreve.47.992} {\bibfield
      {journal} {\bibinfo  {journal} {Physical Review E}\ }\textbf {\bibinfo
      {volume} {47}},\ \bibinfo {pages} {992} (\bibinfo {year} {1993})}\BibitemShut
      {NoStop}%
    \bibitem [{\citenamefont {Vajna}\ \emph {et~al.}(2013)\citenamefont {Vajna},
      \citenamefont {T{\'o}th},\ and\ \citenamefont
      {Kert{\'e}sz}}]{Vajna2013Modelling}%
      \BibitemOpen
      \bibfield  {author} {\bibinfo {author} {\bibfnamefont {S.}~\bibnamefont
      {Vajna}}, \bibinfo {author} {\bibfnamefont {B.}~\bibnamefont {T{\'o}th}},\
      and\ \bibinfo {author} {\bibfnamefont {J.}~\bibnamefont {Kert{\'e}sz}},\
      }\href {https://doi.org/10.1088/1367-2630/15/10/103023} {\bibfield  {journal}
      {\bibinfo  {journal} {New Journal of Physics}\ }\textbf {\bibinfo {volume}
      {15}},\ \bibinfo {pages} {103023} (\bibinfo {year} {2013})}\BibitemShut
      {NoStop}%
    \bibitem [{\citenamefont {Abe}\ and\ \citenamefont
      {Suzuki}(2009)}]{Abe2009Violation}%
      \BibitemOpen
      \bibfield  {author} {\bibinfo {author} {\bibfnamefont {S.}~\bibnamefont
      {Abe}}\ and\ \bibinfo {author} {\bibfnamefont {N.}~\bibnamefont {Suzuki}},\
      }\href {https://doi.org/10.1016/j.physa.2009.01.031} {\bibfield  {journal}
      {\bibinfo  {journal} {Physica A: Statistical Mechanics and its Applications}\
      }\textbf {\bibinfo {volume} {388}},\ \bibinfo {pages} {1917} (\bibinfo {year}
      {2009})}\BibitemShut {NoStop}%
    \bibitem [{\citenamefont {Lee}\ \emph {et~al.}(2018)\citenamefont {Lee},
      \citenamefont {Jung},\ and\ \citenamefont {Jo}}]{Lee2018Hierarchical}%
      \BibitemOpen
      \bibfield  {author} {\bibinfo {author} {\bibfnamefont {B.-H.}\ \bibnamefont
      {Lee}}, \bibinfo {author} {\bibfnamefont {W.-S.}\ \bibnamefont {Jung}},\ and\
      \bibinfo {author} {\bibfnamefont {H.-H.}\ \bibnamefont {Jo}},\ }\href
      {https://doi.org/10.1103/physreve.98.022316} {\bibfield  {journal} {\bibinfo
      {journal} {Physical Review E}\ }\textbf {\bibinfo {volume} {98}},\ \bibinfo
      {pages} {022316} (\bibinfo {year} {2018})}\BibitemShut {NoStop}%
    \bibitem [{\citenamefont {Kim}\ and\ \citenamefont
      {Jo}(2016)}]{Kim2016Measuring}%
      \BibitemOpen
      \bibfield  {author} {\bibinfo {author} {\bibfnamefont {E.-K.}\ \bibnamefont
      {Kim}}\ and\ \bibinfo {author} {\bibfnamefont {H.-H.}\ \bibnamefont {Jo}},\
      }\href {https://doi.org/10.1103/physreve.94.032311} {\bibfield  {journal}
      {\bibinfo  {journal} {Physical Review E}\ }\textbf {\bibinfo {volume} {94}},\
      \bibinfo {pages} {032311} (\bibinfo {year} {2016})}\BibitemShut {NoStop}%
    \bibitem [{\citenamefont {Wang}\ \emph {et~al.}(2015)\citenamefont {Wang},
      \citenamefont {Yuan}, \citenamefont {Pan}, \citenamefont {Jiao},
      \citenamefont {Dai}, \citenamefont {Xue},\ and\ \citenamefont
      {Liu}}]{Wang2015Temporal}%
      \BibitemOpen
      \bibfield  {author} {\bibinfo {author} {\bibfnamefont {W.}~\bibnamefont
      {Wang}}, \bibinfo {author} {\bibfnamefont {N.}~\bibnamefont {Yuan}}, \bibinfo
      {author} {\bibfnamefont {L.}~\bibnamefont {Pan}}, \bibinfo {author}
      {\bibfnamefont {P.}~\bibnamefont {Jiao}}, \bibinfo {author} {\bibfnamefont
      {W.}~\bibnamefont {Dai}}, \bibinfo {author} {\bibfnamefont {G.}~\bibnamefont
      {Xue}},\ and\ \bibinfo {author} {\bibfnamefont {D.}~\bibnamefont {Liu}},\
      }\href {https://doi.org/10.1016/j.physa.2015.05.028} {\bibfield  {journal}
      {\bibinfo  {journal} {Physica A: Statistical Mechanics and its Applications}\
      }\textbf {\bibinfo {volume} {436}},\ \bibinfo {pages} {846} (\bibinfo {year}
      {2015})}\BibitemShut {NoStop}%
    \bibitem [{\citenamefont {Guo}\ \emph {et~al.}(2017)\citenamefont {Guo},
      \citenamefont {Yang}, \citenamefont {Yang}, \citenamefont {Zhao},\ and\
      \citenamefont {Zhou}}]{Guo2017Bounds}%
      \BibitemOpen
      \bibfield  {author} {\bibinfo {author} {\bibfnamefont {F.}~\bibnamefont
      {Guo}}, \bibinfo {author} {\bibfnamefont {D.}~\bibnamefont {Yang}}, \bibinfo
      {author} {\bibfnamefont {Z.}~\bibnamefont {Yang}}, \bibinfo {author}
      {\bibfnamefont {Z.-D.}\ \bibnamefont {Zhao}},\ and\ \bibinfo {author}
      {\bibfnamefont {T.}~\bibnamefont {Zhou}},\ }\href
      {https://doi.org/10.1103/physreve.95.052314} {\bibfield  {journal} {\bibinfo
      {journal} {Physical Review E}\ }\textbf {\bibinfo {volume} {95}},\ \bibinfo
      {pages} {052314} (\bibinfo {year} {2017})}\BibitemShut {NoStop}%
    \bibitem [{\citenamefont {B{\"o}ttcher}\ \emph {et~al.}(2017)\citenamefont
      {B{\"o}ttcher}, \citenamefont {Woolley-Meza},\ and\ \citenamefont
      {Brockmann}}]{Bottcher2017Temporal}%
      \BibitemOpen
      \bibfield  {author} {\bibinfo {author} {\bibfnamefont {L.}~\bibnamefont
      {B{\"o}ttcher}}, \bibinfo {author} {\bibfnamefont {O.}~\bibnamefont
      {Woolley-Meza}},\ and\ \bibinfo {author} {\bibfnamefont {D.}~\bibnamefont
      {Brockmann}},\ }\href {https://doi.org/10.1371/journal.pone.0178062}
      {\bibfield  {journal} {\bibinfo  {journal} {PLoS ONE}\ }\textbf {\bibinfo
      {volume} {12}},\ \bibinfo {pages} {e0178062} (\bibinfo {year}
      {2017})}\BibitemShut {NoStop}%
    \bibitem [{\citenamefont {Karsai}\ \emph
      {et~al.}(2012{\natexlab{b}})\citenamefont {Karsai}, \citenamefont {Kaski},\
      and\ \citenamefont {Kert{\'e}sz}}]{Karsai2012Correlated}%
      \BibitemOpen
      \bibfield  {author} {\bibinfo {author} {\bibfnamefont {M.}~\bibnamefont
      {Karsai}}, \bibinfo {author} {\bibfnamefont {K.}~\bibnamefont {Kaski}},\ and\
      \bibinfo {author} {\bibfnamefont {J.}~\bibnamefont {Kert{\'e}sz}},\ }\href
      {https://doi.org/10.1371/journal.pone.0040612} {\bibfield  {journal}
      {\bibinfo  {journal} {PLoS ONE}\ }\textbf {\bibinfo {volume} {7}},\ \bibinfo
      {pages} {e40612} (\bibinfo {year} {2012}{\natexlab{b}})}\BibitemShut
      {NoStop}%
    \bibitem [{\citenamefont {Yasseri}\ \emph {et~al.}(2012)\citenamefont
      {Yasseri}, \citenamefont {Sumi}, \citenamefont {Rung}, \citenamefont
      {Kornai},\ and\ \citenamefont {Kert{\'e}sz}}]{Yasseri2012Dynamics}%
      \BibitemOpen
      \bibfield  {author} {\bibinfo {author} {\bibfnamefont {T.}~\bibnamefont
      {Yasseri}}, \bibinfo {author} {\bibfnamefont {R.}~\bibnamefont {Sumi}},
      \bibinfo {author} {\bibfnamefont {A.}~\bibnamefont {Rung}}, \bibinfo {author}
      {\bibfnamefont {A.}~\bibnamefont {Kornai}},\ and\ \bibinfo {author}
      {\bibfnamefont {J.}~\bibnamefont {Kert{\'e}sz}},\ }\href
      {https://doi.org/10.1371/journal.pone.0038869} {\bibfield  {journal}
      {\bibinfo  {journal} {PLoS ONE}\ }\textbf {\bibinfo {volume} {7}},\ \bibinfo
      {pages} {e38869} (\bibinfo {year} {2012})}\BibitemShut {NoStop}%
    \bibitem [{Note1()}]{Note1}%
      \BibitemOpen
      \bibinfo {note} {The generalized memory coefficient has also been suggested
      as the Pearson correlation coefficient between two IETs separated by $k$
      IETs~\cite {Goh2008Burstiness}. The case with $k=0$ corresponds to the $M$ in
      Eq.~(\ref {eq:memory_original}). The relation between the generalized memory
      coefficients and burst size distributions can be studied for better
      understanding the correlation structure between IETs.}\BibitemShut {Stop}%
    \bibitem [{\citenamefont {Jiang}\ \emph {et~al.}(2016)\citenamefont {Jiang},
      \citenamefont {Xie}, \citenamefont {Li}, \citenamefont {Zhou},\ and\
      \citenamefont {Sornette}}]{Jiang2016Twostate}%
      \BibitemOpen
      \bibfield  {author} {\bibinfo {author} {\bibfnamefont {Z.-Q.}\ \bibnamefont
      {Jiang}}, \bibinfo {author} {\bibfnamefont {W.-J.}\ \bibnamefont {Xie}},
      \bibinfo {author} {\bibfnamefont {M.-X.}\ \bibnamefont {Li}}, \bibinfo
      {author} {\bibfnamefont {W.-X.}\ \bibnamefont {Zhou}},\ and\ \bibinfo
      {author} {\bibfnamefont {D.}~\bibnamefont {Sornette}},\ }\href
      {https://doi.org/10.1088/1742-5468/2016/07/073210} {\bibfield  {journal}
      {\bibinfo  {journal} {Journal of Statistical Mechanics: Theory and
      Experiment}\ }\textbf {\bibinfo {volume} {2016}},\ \bibinfo {pages} {073210}
      (\bibinfo {year} {2016})}\BibitemShut {NoStop}%
    \bibitem [{\citenamefont {Jo}\ \emph {et~al.}(2015)\citenamefont {Jo},
      \citenamefont {Perotti}, \citenamefont {Kaski},\ and\ \citenamefont
      {Kert{\'e}sz}}]{Jo2015Correlated}%
      \BibitemOpen
      \bibfield  {author} {\bibinfo {author} {\bibfnamefont {H.-H.}\ \bibnamefont
      {Jo}}, \bibinfo {author} {\bibfnamefont {J.~I.}\ \bibnamefont {Perotti}},
      \bibinfo {author} {\bibfnamefont {K.}~\bibnamefont {Kaski}},\ and\ \bibinfo
      {author} {\bibfnamefont {J.}~\bibnamefont {Kert{\'e}sz}},\ }\href
      {https://doi.org/10.1103/physreve.92.022814} {\bibfield  {journal} {\bibinfo
      {journal} {Physical Review E}\ }\textbf {\bibinfo {volume} {92}},\ \bibinfo
      {pages} {022814} (\bibinfo {year} {2015})}\BibitemShut {NoStop}%
    \bibitem [{\citenamefont {Jo}(2019)}]{Jo2019Analytically}%
      \BibitemOpen
      \bibfield  {author} {\bibinfo {author} {\bibfnamefont {H.-H.}\ \bibnamefont
      {Jo}},\ }\href {https://doi.org/10.1103/PhysRevE.100.012306} {\bibfield
      {journal} {\bibinfo  {journal} {Physical Review E}\ }\textbf {\bibinfo
      {volume} {100}},\ \bibinfo {pages} {012306} (\bibinfo {year}
      {2019})}\BibitemShut {NoStop}%
    \bibitem [{\citenamefont {Jo}\ and\ \citenamefont
      {Hiraoka}(2018)}]{Jo2018Limits}%
      \BibitemOpen
      \bibfield  {author} {\bibinfo {author} {\bibfnamefont {H.-H.}\ \bibnamefont
      {Jo}}\ and\ \bibinfo {author} {\bibfnamefont {T.}~\bibnamefont {Hiraoka}},\
      }\href {https://doi.org/10.1103/physreve.97.032121} {\bibfield  {journal}
      {\bibinfo  {journal} {Physical Review E}\ }\textbf {\bibinfo {volume} {97}},\
      \bibinfo {pages} {032121} (\bibinfo {year} {2018})}\BibitemShut {NoStop}%
    \bibitem [{\citenamefont {Vazquez}(2007)}]{Vazquez2007Impact}%
      \BibitemOpen
      \bibfield  {author} {\bibinfo {author} {\bibfnamefont {A.}~\bibnamefont
      {Vazquez}},\ }\href {https://doi.org/10.1016/j.physa.2006.04.060} {\bibfield
      {journal} {\bibinfo  {journal} {Physica A: Statistical Mechanics and its
      Applications}\ }\textbf {\bibinfo {volume} {373}},\ \bibinfo {pages} {747}
      (\bibinfo {year} {2007})}\BibitemShut {NoStop}%
    \bibitem [{\citenamefont {Karsai}\ \emph {et~al.}(2011)\citenamefont {Karsai},
      \citenamefont {Kivel{\"a}}, \citenamefont {Pan}, \citenamefont {Kaski},
      \citenamefont {Kert{\'e}sz}, \citenamefont {Barab{\'a}si},\ and\
      \citenamefont {Saram{\"a}ki}}]{Karsai2011Small}%
      \BibitemOpen
      \bibfield  {author} {\bibinfo {author} {\bibfnamefont {M.}~\bibnamefont
      {Karsai}}, \bibinfo {author} {\bibfnamefont {M.}~\bibnamefont {Kivel{\"a}}},
      \bibinfo {author} {\bibfnamefont {R.~K.}\ \bibnamefont {Pan}}, \bibinfo
      {author} {\bibfnamefont {K.}~\bibnamefont {Kaski}}, \bibinfo {author}
      {\bibfnamefont {J.}~\bibnamefont {Kert{\'e}sz}}, \bibinfo {author}
      {\bibfnamefont {A.-L.}\ \bibnamefont {Barab{\'a}si}},\ and\ \bibinfo {author}
      {\bibfnamefont {J.}~\bibnamefont {Saram{\"a}ki}},\ }\href
      {https://doi.org/10.1103/physreve.83.025102} {\bibfield  {journal} {\bibinfo
      {journal} {Physical Review E}\ }\textbf {\bibinfo {volume} {83}},\ \bibinfo
      {pages} {025102(R)} (\bibinfo {year} {2011})}\BibitemShut {NoStop}%
    \bibitem [{\citenamefont {Miritello}\ \emph {et~al.}(2011)\citenamefont
      {Miritello}, \citenamefont {Moro},\ and\ \citenamefont
      {Lara}}]{Miritello2011Dynamical}%
      \BibitemOpen
      \bibfield  {author} {\bibinfo {author} {\bibfnamefont {G.}~\bibnamefont
      {Miritello}}, \bibinfo {author} {\bibfnamefont {E.}~\bibnamefont {Moro}},\
      and\ \bibinfo {author} {\bibfnamefont {R.}~\bibnamefont {Lara}},\ }\href
      {https://doi.org/10.1103/physreve.83.045102} {\bibfield  {journal} {\bibinfo
      {journal} {Physical Review E}\ }\textbf {\bibinfo {volume} {83}},\ \bibinfo
      {pages} {045102(R)} (\bibinfo {year} {2011})}\BibitemShut {NoStop}%
    \bibitem [{\citenamefont {Iribarren}\ and\ \citenamefont
      {Moro}(2009)}]{Iribarren2009Impact}%
      \BibitemOpen
      \bibfield  {author} {\bibinfo {author} {\bibfnamefont {J.~L.}\ \bibnamefont
      {Iribarren}}\ and\ \bibinfo {author} {\bibfnamefont {E.}~\bibnamefont
      {Moro}},\ }\href {https://doi.org/10.1103/physrevlett.103.038702} {\bibfield
      {journal} {\bibinfo  {journal} {Physical Review Letters}\ }\textbf {\bibinfo
      {volume} {103}},\ \bibinfo {pages} {038702} (\bibinfo {year}
      {2009})}\BibitemShut {NoStop}%
    \bibitem [{\citenamefont {Rocha}\ \emph {et~al.}(2011)\citenamefont {Rocha},
      \citenamefont {Liljeros},\ and\ \citenamefont {Holme}}]{Rocha2011Simulated}%
      \BibitemOpen
      \bibfield  {author} {\bibinfo {author} {\bibfnamefont {L.~E.~C.}\
      \bibnamefont {Rocha}}, \bibinfo {author} {\bibfnamefont {F.}~\bibnamefont
      {Liljeros}},\ and\ \bibinfo {author} {\bibfnamefont {P.}~\bibnamefont
      {Holme}},\ }\href {https://doi.org/10.1371/journal.pcbi.1001109} {\bibfield
      {journal} {\bibinfo  {journal} {PLOS Computational Biology}\ }\textbf
      {\bibinfo {volume} {7}},\ \bibinfo {pages} {e1001109} (\bibinfo {year}
      {2011})}\BibitemShut {NoStop}%
    \bibitem [{\citenamefont {Rocha}\ and\ \citenamefont
      {Blondel}(2013)}]{Rocha2013Bursts}%
      \BibitemOpen
      \bibfield  {author} {\bibinfo {author} {\bibfnamefont {L.~E.~C.}\
      \bibnamefont {Rocha}}\ and\ \bibinfo {author} {\bibfnamefont {V.~D.}\
      \bibnamefont {Blondel}},\ }\href
      {https://doi.org/10.1371/journal.pcbi.1002974} {\bibfield  {journal}
      {\bibinfo  {journal} {PLOS Computational Biology}\ }\textbf {\bibinfo
      {volume} {9}},\ \bibinfo {pages} {e1002974} (\bibinfo {year}
      {2013})}\BibitemShut {NoStop}%
    \bibitem [{\citenamefont {Takaguchi}\ \emph {et~al.}(2013)\citenamefont
      {Takaguchi}, \citenamefont {Masuda},\ and\ \citenamefont
      {Holme}}]{Takaguchi2013Bursty}%
      \BibitemOpen
      \bibfield  {author} {\bibinfo {author} {\bibfnamefont {T.}~\bibnamefont
      {Takaguchi}}, \bibinfo {author} {\bibfnamefont {N.}~\bibnamefont {Masuda}},\
      and\ \bibinfo {author} {\bibfnamefont {P.}~\bibnamefont {Holme}},\ }\href
      {https://doi.org/10.1371/journal.pone.0068629} {\bibfield  {journal}
      {\bibinfo  {journal} {PLoS ONE}\ }\textbf {\bibinfo {volume} {8}},\ \bibinfo
      {pages} {e68629} (\bibinfo {year} {2013})}\BibitemShut {NoStop}%
    \bibitem [{\citenamefont {Masuda}\ and\ \citenamefont
      {Holme}(2013)}]{Masuda2013Predicting}%
      \BibitemOpen
      \bibfield  {author} {\bibinfo {author} {\bibfnamefont {N.}~\bibnamefont
      {Masuda}}\ and\ \bibinfo {author} {\bibfnamefont {P.}~\bibnamefont {Holme}},\
      }\href {https://doi.org/10.12703/p5-6} {\bibfield  {journal} {\bibinfo
      {journal} {F1000Prime Reports}\ }\textbf {\bibinfo {volume} {5}},\ \bibinfo
      {pages} {6} (\bibinfo {year} {2013})}\BibitemShut {NoStop}%
    \bibitem [{\citenamefont {Jo}\ \emph {et~al.}(2014)\citenamefont {Jo},
      \citenamefont {Perotti}, \citenamefont {Kaski},\ and\ \citenamefont
      {Kert{\'e}sz}}]{Jo2014Analytically}%
      \BibitemOpen
      \bibfield  {author} {\bibinfo {author} {\bibfnamefont {H.-H.}\ \bibnamefont
      {Jo}}, \bibinfo {author} {\bibfnamefont {J.~I.}\ \bibnamefont {Perotti}},
      \bibinfo {author} {\bibfnamefont {K.}~\bibnamefont {Kaski}},\ and\ \bibinfo
      {author} {\bibfnamefont {J.}~\bibnamefont {Kert{\'e}sz}},\ }\href
      {https://doi.org/10.1103/physrevx.4.011041} {\bibfield  {journal} {\bibinfo
      {journal} {Physical Review X}\ }\textbf {\bibinfo {volume} {4}},\ \bibinfo
      {pages} {011041} (\bibinfo {year} {2014})}\BibitemShut {NoStop}%
    \bibitem [{\citenamefont {Perotti}\ \emph {et~al.}(2014)\citenamefont
      {Perotti}, \citenamefont {Jo}, \citenamefont {Holme},\ and\ \citenamefont
      {Saram{\"a}ki}}]{Perotti2014Temporal}%
      \BibitemOpen
      \bibfield  {author} {\bibinfo {author} {\bibfnamefont {J.~I.}\ \bibnamefont
      {Perotti}}, \bibinfo {author} {\bibfnamefont {H.-H.}\ \bibnamefont {Jo}},
      \bibinfo {author} {\bibfnamefont {P.}~\bibnamefont {Holme}},\ and\ \bibinfo
      {author} {\bibfnamefont {J.}~\bibnamefont {Saram{\"a}ki}}} (\bibinfo {year}
      {2014}),\ \bibinfo {note} {arXiv:1411.5553}\BibitemShut {NoStop}%
    \bibitem [{\citenamefont {Delvenne}\ \emph {et~al.}(2015)\citenamefont
      {Delvenne}, \citenamefont {Lambiotte},\ and\ \citenamefont
      {Rocha}}]{Delvenne2015Diffusion}%
      \BibitemOpen
      \bibfield  {author} {\bibinfo {author} {\bibfnamefont {J.-C.}\ \bibnamefont
      {Delvenne}}, \bibinfo {author} {\bibfnamefont {R.}~\bibnamefont
      {Lambiotte}},\ and\ \bibinfo {author} {\bibfnamefont {L.~E.~C.}\ \bibnamefont
      {Rocha}},\ }\href {https://doi.org/10.1038/ncomms8366} {\bibfield  {journal}
      {\bibinfo  {journal} {Nature Communications}\ }\textbf {\bibinfo {volume}
      {6}},\ \bibinfo {pages} {7366} (\bibinfo {year} {2015})}\BibitemShut
      {NoStop}%
    \bibitem [{\citenamefont {Pastor-Satorras}\ \emph {et~al.}(2015)\citenamefont
      {Pastor-Satorras}, \citenamefont {Castellano}, \citenamefont {Van~Mieghem},\
      and\ \citenamefont {Vespignani}}]{Pastor-Satorras2015Epidemic}%
      \BibitemOpen
      \bibfield  {author} {\bibinfo {author} {\bibfnamefont {R.}~\bibnamefont
      {Pastor-Satorras}}, \bibinfo {author} {\bibfnamefont {C.}~\bibnamefont
      {Castellano}}, \bibinfo {author} {\bibfnamefont {P.}~\bibnamefont
      {Van~Mieghem}},\ and\ \bibinfo {author} {\bibfnamefont {A.}~\bibnamefont
      {Vespignani}},\ }\href {https://doi.org/10.1103/RevModPhys.87.925} {\bibfield
       {journal} {\bibinfo  {journal} {Reviews of Modern Physics}\ }\textbf
      {\bibinfo {volume} {87}},\ \bibinfo {pages} {925} (\bibinfo {year}
      {2015})}\BibitemShut {NoStop}%
    \bibitem [{\citenamefont {Artime}\ \emph {et~al.}(2017)\citenamefont {Artime},
      \citenamefont {Ramasco},\ and\ \citenamefont
      {San~Miguel}}]{Artime2017Dynamics}%
      \BibitemOpen
      \bibfield  {author} {\bibinfo {author} {\bibfnamefont {O.}~\bibnamefont
      {Artime}}, \bibinfo {author} {\bibfnamefont {J.~J.}\ \bibnamefont
      {Ramasco}},\ and\ \bibinfo {author} {\bibfnamefont {M.}~\bibnamefont
      {San~Miguel}},\ }\href {https://doi.org/10.1038/srep41627} {\bibfield
      {journal} {\bibinfo  {journal} {Scientific Reports}\ }\textbf {\bibinfo
      {volume} {7}},\ \bibinfo {pages} {41627} (\bibinfo {year}
      {2017})}\BibitemShut {NoStop}%
    \bibitem [{\citenamefont {Hiraoka}\ and\ \citenamefont
      {Jo}(2018)}]{Hiraoka2018Correlated}%
      \BibitemOpen
      \bibfield  {author} {\bibinfo {author} {\bibfnamefont {T.}~\bibnamefont
      {Hiraoka}}\ and\ \bibinfo {author} {\bibfnamefont {H.-H.}\ \bibnamefont
      {Jo}},\ }\href {https://doi.org/10.1038/s41598-018-33700-8} {\bibfield
      {journal} {\bibinfo  {journal} {Scientific Reports}\ }\textbf {\bibinfo
      {volume} {8}},\ \bibinfo {pages} {15321} (\bibinfo {year}
      {2018})}\BibitemShut {NoStop}%
    \bibitem [{\citenamefont {Masuda}\ and\ \citenamefont
      {Rocha}(2018)}]{Masuda2018Gillespie}%
      \BibitemOpen
      \bibfield  {author} {\bibinfo {author} {\bibfnamefont {N.}~\bibnamefont
      {Masuda}}\ and\ \bibinfo {author} {\bibfnamefont {L.~E.~C.}\ \bibnamefont
      {Rocha}},\ }\href {https://doi.org/10.1137/16m1055876} {\bibfield  {journal}
      {\bibinfo  {journal} {SIAM Review}\ }\textbf {\bibinfo {volume} {60}},\
      \bibinfo {pages} {95} (\bibinfo {year} {2018})}\BibitemShut {NoStop}%
    \bibitem [{\citenamefont {Gueuning}\ \emph {et~al.}(2015)\citenamefont
      {Gueuning}, \citenamefont {Delvenne},\ and\ \citenamefont
      {Lambiotte}}]{Gueuning2015Imperfect}%
      \BibitemOpen
      \bibfield  {author} {\bibinfo {author} {\bibfnamefont {M.}~\bibnamefont
      {Gueuning}}, \bibinfo {author} {\bibfnamefont {J.-C.}\ \bibnamefont
      {Delvenne}},\ and\ \bibinfo {author} {\bibfnamefont {R.}~\bibnamefont
      {Lambiotte}},\ }\href {https://doi.org/10.1140/epjb%252fe2015-60596-0}
      {\bibfield  {journal} {\bibinfo  {journal} {The European Physical Journal B}\
      }\textbf {\bibinfo {volume} {88}},\ \bibinfo {pages} {282} (\bibinfo {year}
      {2015})}\BibitemShut {NoStop}%
    \bibitem [{\citenamefont {Janssen}\ \emph {et~al.}(2004)\citenamefont
      {Janssen}, \citenamefont {M{\"u}ller},\ and\ \citenamefont
      {Stenull}}]{Janssen2004Generalized}%
      \BibitemOpen
      \bibfield  {author} {\bibinfo {author} {\bibfnamefont {H.-K.}\ \bibnamefont
      {Janssen}}, \bibinfo {author} {\bibfnamefont {M.}~\bibnamefont
      {M{\"u}ller}},\ and\ \bibinfo {author} {\bibfnamefont {O.}~\bibnamefont
      {Stenull}},\ }\href {https://doi.org/10.1103/physreve.70.026114} {\bibfield
      {journal} {\bibinfo  {journal} {Physical Review E}\ }\textbf {\bibinfo
      {volume} {70}},\ \bibinfo {pages} {026114} (\bibinfo {year}
      {2004})}\BibitemShut {NoStop}%
    \bibitem [{\citenamefont {Dodds}\ and\ \citenamefont
      {Watts}(2004)}]{Dodds2004Universal}%
      \BibitemOpen
      \bibfield  {author} {\bibinfo {author} {\bibfnamefont {P.~S.}\ \bibnamefont
      {Dodds}}\ and\ \bibinfo {author} {\bibfnamefont {D.~J.}\ \bibnamefont
      {Watts}},\ }\href {https://doi.org/10.1103/physrevlett.92.218701} {\bibfield
      {journal} {\bibinfo  {journal} {Physical Review Letters}\ }\textbf {\bibinfo
      {volume} {92}},\ \bibinfo {pages} {218701} (\bibinfo {year}
      {2004})}\BibitemShut {NoStop}%
    \bibitem [{\citenamefont {Bizhani}\ \emph {et~al.}(2012)\citenamefont
      {Bizhani}, \citenamefont {Paczuski},\ and\ \citenamefont
      {Grassberger}}]{Bizhani2012Discontinuous}%
      \BibitemOpen
      \bibfield  {author} {\bibinfo {author} {\bibfnamefont {G.}~\bibnamefont
      {Bizhani}}, \bibinfo {author} {\bibfnamefont {M.}~\bibnamefont {Paczuski}},\
      and\ \bibinfo {author} {\bibfnamefont {P.}~\bibnamefont {Grassberger}},\
      }\href {https://doi.org/10.1103/physreve.86.011128} {\bibfield  {journal}
      {\bibinfo  {journal} {Physical Review E}\ }\textbf {\bibinfo {volume} {86}},\
      \bibinfo {pages} {011128} (\bibinfo {year} {2012})}\BibitemShut {NoStop}%
    \bibitem [{\citenamefont {Chung}\ \emph {et~al.}(2014)\citenamefont {Chung},
      \citenamefont {Baek}, \citenamefont {Kim}, \citenamefont {Ha},\ and\
      \citenamefont {Jeong}}]{Chung2014Generalized}%
      \BibitemOpen
      \bibfield  {author} {\bibinfo {author} {\bibfnamefont {K.}~\bibnamefont
      {Chung}}, \bibinfo {author} {\bibfnamefont {Y.}~\bibnamefont {Baek}},
      \bibinfo {author} {\bibfnamefont {D.}~\bibnamefont {Kim}}, \bibinfo {author}
      {\bibfnamefont {M.}~\bibnamefont {Ha}},\ and\ \bibinfo {author}
      {\bibfnamefont {H.}~\bibnamefont {Jeong}},\ }\href
      {https://doi.org/10.1103/physreve.89.052811} {\bibfield  {journal} {\bibinfo
      {journal} {Physical Review E}\ }\textbf {\bibinfo {volume} {89}},\ \bibinfo
      {pages} {052811} (\bibinfo {year} {2014})}\BibitemShut {NoStop}%
    \bibitem [{Note2()}]{Note2}%
      \BibitemOpen
      \bibinfo {note} {Another algorithm for generating bursty time series using
      the copula has recently been suggested~\cite
      {Jo2019Copulabased}.}\BibitemShut {Stop}%
    \bibitem [{\citenamefont {Kimmel}\ and\ \citenamefont
      {Axelrod}(2002)}]{Kimmel2002Branching}%
      \BibitemOpen
      \bibfield  {author} {\bibinfo {author} {\bibfnamefont {M.}~\bibnamefont
      {Kimmel}}\ and\ \bibinfo {author} {\bibfnamefont {D.~E.}\ \bibnamefont
      {Axelrod}},\ }\href {https://doi.org/10.1007/b97371} {\emph {\bibinfo {title}
      {Branching {Processes} in {Biology}}}},\ Vol.~\bibinfo {volume} {19}\
      (\bibinfo  {publisher} {Springer New York},\ \bibinfo {address} {New York,
      NY},\ \bibinfo {year} {2002})\BibitemShut {NoStop}%
    \bibitem [{\citenamefont {Jo}\ \emph {et~al.}(2013)\citenamefont {Jo},
      \citenamefont {Pan}, \citenamefont {Perotti},\ and\ \citenamefont
      {Kaski}}]{Jo2013Contextual}%
      \BibitemOpen
      \bibfield  {author} {\bibinfo {author} {\bibfnamefont {H.-H.}\ \bibnamefont
      {Jo}}, \bibinfo {author} {\bibfnamefont {R.~K.}\ \bibnamefont {Pan}},
      \bibinfo {author} {\bibfnamefont {J.~I.}\ \bibnamefont {Perotti}},\ and\
      \bibinfo {author} {\bibfnamefont {K.}~\bibnamefont {Kaski}},\ }\href
      {https://doi.org/10.1103/physreve.87.062131} {\bibfield  {journal} {\bibinfo
      {journal} {Physical Review E}\ }\textbf {\bibinfo {volume} {87}},\ \bibinfo
      {pages} {062131} (\bibinfo {year} {2013})}\BibitemShut {NoStop}%
    \bibitem [{\citenamefont {Jo}\ \emph {et~al.}(2019)\citenamefont {Jo},
      \citenamefont {Lee}, \citenamefont {Hiraoka},\ and\ \citenamefont
      {Jung}}]{Jo2019Copulabased}%
      \BibitemOpen
      \bibfield  {author} {\bibinfo {author} {\bibfnamefont {H.-H.}\ \bibnamefont
      {Jo}}, \bibinfo {author} {\bibfnamefont {B.-H.}\ \bibnamefont {Lee}},
      \bibinfo {author} {\bibfnamefont {T.}~\bibnamefont {Hiraoka}},\ and\ \bibinfo
      {author} {\bibfnamefont {W.-S.}\ \bibnamefont {Jung}}} (\bibinfo {year}
      {2019}),\ \bibinfo {note} {arXiv:1904.08795}\BibitemShut {NoStop}%
    \end{thebibliography}
%
    
\end{document}